\documentclass[fleqn,usenatbib]{mnras}

\usepackage{newtxtext,newtxmath}

\usepackage[T1]{fontenc}

\DeclareRobustCommand{\VAN}[3]{#2}
\let\VANthebibliography\thebibliography
\def\thebibliography{\DeclareRobustCommand{\VAN}[3]{##3}\VANthebibliography}

\usepackage{url}
\usepackage[utf8]{inputenc}
\usepackage{graphicx}
\usepackage{amsmath}
\usepackage{siunitx}
\usepackage{natbib}
\usepackage{physics}
\usepackage{chemformula}
\usepackage{gensymb}
\usepackage{float}

\usepackage{graphicx}	
\usepackage{amsmath}	

\newcommand{\fidloc}{\SI{1}{au}}	

\title[Radiative torque impact on heliospheric dust]{Investigating the influence of the radiative torque disruption on the size evolution of dust in the heliosphere}

\author[Chi-Hang Ng et al.]{
Chi-Hang Ng,$^{1,2}$ 
Pin-Gao Gu$^{2}$\thanks{E-mail: gu@asiaa.sinica.edu.tw}
and Thiem Hoang$^{3,4}$
\\
$^{1}$Department of Physics, National Taiwan University, 106319 Taipei, Taiwan\\
$^{2}$Institute of Astronomy and Astrophysics, Academia Sinica, 106319 Taipei, Taiwan\\
$^{3}$Korea Astronomy and Space Science Institute, Daejeon 34055, Republic of Korea\\
$^{4}$ Department of Astronomy and Space Science, University of Science and Technology, 217 Gajeong-ro, Yuseong-gu, Daejeon, 34113, Republic of Korea
}

\date{Accepted XXX. Received YYY; in original form ZZZ}

\pubyear{2015}

\begin{document}
\label{firstpage}
\pagerange{\pageref{firstpage}--\pageref{lastpage}}
\maketitle

\begin{abstract}
In this paper, we conduct a detailed study on the effect of Radiative Torque Disruption (RATD) mechanism on the fragmentation of micrometer-sized dust grains into nanoparticles within the heliosphere. We start by estimating the disruption timescales for dust grains under various centrifugal stresses. Our numerical calculations demonstrate that RATD is a highly effective mechanism for breaking down micrometer-sized grains, producing nanoparticles more efficiently than other fragmentation processes. RATD also prevents micrometer-sized grains from being expelled by radiation pressure. Our findings indicate that the location of the present water snow line depends not only on temperature but also on the size of dust grains. For smaller grains, the snow line can shift outward beyond the position defined by thermal sublimation. Furthermore, we model the size distribution of dust grains modified by the RATD mechanism using a simplified model, showing that rotational disruption significantly decreases the number density of micrometer-sized grains while substantially increasing the number density of sub-micrometer-sized grains. However, the fraction of dust grains aligned at high-$J$ attractors by radiative torques less than 80\% can considerably weaken the effect of RATD on the grain size distribution. Finally, we suggest several experiments that could potentially test the RATD mechanism and discuss the uncertainties of our model in more realistic applications to heliospheric dust.
\end{abstract}
\begin{keywords}
dust, extinction -- ISM: evolution -- ISM: general -- solar wind
\end{keywords}



\section{Introduction}

Remote sensing and in situ measurements performed in the heliosphere of the Solar System showed that very small dust grains (radius below tens of nanometers, hereafter nanoparticles) in the Solar System can be produced by various mechanisms of fragmentation. The first in situ measurement of micrometer size dust dates back to the Voyager missions and their visits to Saturn, Uranus, and Neptune \citep{OShea2017}. An intense unpolarized broadband noise was detected by the planetary radio astronomy experiment during the experiment on Voyager 2, probably due to dust impacts on the spacecraft \citep{Pedersen1991}. Later, dust particles were probed by the Ulysses and Cassini spacecrafts \citep[e.g.,][]{Mann2017}. The Ulysses mission was intended to provide direct observations of dust grains with an in situ dust detector on board the Ulysses spacecraft \citep{1992A&AS...92..411G,Kruger2015}. These in-situ measurements successfully detected interstellar dust (ISD), interplanetary dust particle (IDP), and circumplanetary dust \citep[e.g.,][]{jorgensen_distribution_2021,Sterken2019,hsu_situ_2018}. The Cassini spacecraft has detected dust impacts around Saturn using both antennas \citep{Ye2016} and a dedicated dust instrument \citep{Kempf2008}. The observed stream particles found by the Cassini flyby around Jupiter originate from the volcanic active jovian satellite Io \citep{postberg_composition_2006}. The data collected by Cassini spacecraft on its way to Saturn suggested that the nanoparticles are produced in the innermost heliosphere and picked up by solar wind \citep{Schippers2015}. Dust impact signals are also detected by the S/WAVES plasma wave antenna instruments on the STEREO spacecraft \citep{OShea2017}, and a very high velocity of nanoparticles is proposed to explain the frequency of the STEREO A events \citep{Meyer2009}. The high velocity of nanoparticles can be explained by the acceleration of charged particles picked up by solar wind \citep{Czechowski2010}. Nanodust from comet 67P/Churyumov-Gerasimenko has been collected and measured by MIDAS (Micro Imaging Dust Analysis System) \citep{mannel2019dust}, and the IES (ion and electron sensor) onboard the Rosetta spacecraft \citep{llera2020simultaneous}. 

In addition, remote observations show the ubiquity of nanoparticles in debris disks. Observations reveal NIR emission excess at the innermost regions of debris disks, possibly near the silica sublimation zone \citep{Ciardi2001}. Later, \citet{Defrere2011} also found hot dust in Vega. \citet{Su2013} detected the emission excess at $\SI{2}{\mu m}$ toward two nearby debris disks around Vega and Fomalhaut. They pointed out that to reproduce NIR excess at $\sim \SI{2}{\mu m}$ observed toward Vega and Fomalhaut, it requires nanoparticles between $5-20\,\si{.nm}$ located within tens of stellar radii. \citet{VanLieshout2014} suggested that nanoparticles arise from collisions of cometary dust grains \citep[see also][]{Lebreton2013}. Spectroscopic observations in \cite{Bhowmik2019} suggested that very small grains should be present in the debris disk around HD 32297. Spectroscopic data showed that sub-$\mu m$-sized silicas are found toward HD 113766 and HD 172555 \citep{Su2020}. Various studies suggested that the production of nanoparticles can be caused by sublimation and collisional fragmentation \citep{Mann2007,Mann2017,Su2020}. 

\citet{1970BAICz..21..204H} suggested that the absorption and emission of photons could result in the rapid rotation of a dust grain, but the Harwit model is based on the random walk theory of photon angular momentum. Later one, \cite{Dolginov.1976} suggested that the differential absorption and scattering of radiation by helical grains would induce radiative torques and cause grain alignment. Numerical calculations by \citet{Draine1996} showed that RATs are significant for irregular grain shape and can spin up grains to superthermal rotation (i.e., grain angular momentum ($J$) exceeding the thermal angular value). An analytical model for RATs was introduced in \cite{Lazarian2007a}. Detailed numerical calculations in \cite{hoang_radiative_2008} for grain alignment by RATs revealed that grains can be aligned at low-$J$ attractor and high-$J$ attractor points. The fraction of grains aligned at high-$J$, $f_\mathrm{highJ}$, depends on the radiation field, grain size, shape, charge distribution, and magnetic properties \citep{hoang_unified_2016, hoang_internal_2023}. \citet{Hoang2019_3} first recognized that large grains aligned at high-$J$ attractors can be broken into small particles when the centrifugal force induced by grain suprathermal rotation exceeds the binding force that holds the grain components (also \citealt{Hoang2019}). Recently, \citet{hoang_effect_2021} showed that RATD can efficiently produce nanoparticles by fragmenting large particles in the F-corona. The model provided an additional mechanism for the observed decrease of F-coronal dust away from the Sun. Nonetheless, how RATD influences dust fragmentation beyond the F-corona in the heliosphere has not been explored. 

In this paper, we investigate whether nanoparticles can also be produced by the fragmentation of RATD from large particles ($>\SI{0.1}{\mu m}$) in the Solar System environment. Several important processes, such as collisions and radiative effects, have been modeled for dust size evolutions in the interplanetary medium \cite[e.g.,][for a review]{Mann2017}. In this work, we attempt to study dust size distribution by incorporating RATD in a simplified collisional cascade model with Poynting-Robertson drag. We also conduct a rudimentary study to examine the RATD of a $\mu m$-sized particle strongly subject to radiation pressure. Despite simplicity of our modeling approach, this work aims to hint at the potential impact of RATD on heliospheric dust.

The paper is structured as follows. First, we remind the reader of the theory of RATD, which includes the equations of rotational disruption timescales and collisional disruption timescales, and mention the parameters of the heliosphere. Additionally, we introduce a simple model for calculating the size distribution (Sect. \ref{RATD}). Subsequently, we apply the RATD theory to the dust inside the heliosphere and compute the disruption grain size and timescales. The size distribution of the grains is calculated and compared the results with different tensile strength and fraction of grain alignment. We also apply the RATD theory to constrain the location of the present water snow line (Sect. \ref{result}). Sect. \ref{discuss} discusses the effects of radiation pressure and possible future experimental measurements. Finally, we summarize the findings of this paper in Sect. \ref{summary}. Details of the derivation of the general rotational timescale of a dust grain damped by hydrogen gas are given in Appendix \ref{Appendix A}, and the derivation of the radiation pressure cross-section is given in Appendix \ref{Appendix B}. We further introduce and compare the magnitudes of different torques arising from alignment mechanisms in Appendix \ref{Appendix C}.

\section{Theoretical foundation of RATD}
\label{RATD}
In this section, we remind the reader of the theory of RATD \citet{Hoang2019_3}, which has been investigated extensively for different environments and observational testing (see \citealt{TramHoang.2022}). The underlying idea is that, due to the effect of radiative torques, dust grains can be aligned and rotate suprathermally at an attractor point of high angular momentum (hereafter high-$J$ attractors) \citep{Lazarian2007a,hoang_radiative_2008}. Moreover, gas collisions and infrared emission will decrease the grain rotational angular velocity. Consequently, the timescales of different mechanisms of fragmentation are vitally important to the final number density distribution of the dust grains.

\subsection{Radiation Field, Gas and Dust Properties in the Heliosphere}
\label{helio_Properties}
Let $u_{\mathrm{rad}}=\int{u_{\lambdaup}d\lambdaup}$ be the energy density of the radiation field and define $U=u_{\mathrm{rad}}/u_\mathrm{ISRF}$, with the energy density of the average interstellar radiation
field (ISRF) $u_\mathrm{ISRF}=\SI{8.64e-13}{erg.cm^{-3}}$ \citep{mathis1983interstellar}. Therefore, in the environment of the Solar System, $U\approx\num{5.2e7}(\frac{r}{\SI{1}{au}})^{-2}$.

The radiation intensity depends on the surface temperature of the Sun. We assume the emission of the Sun is perfect black body radiation with a surface temperature of \SI{5800}{K}. Therefore, the energy density of the radiation field at the heliocentric distance $r$ is $u_{\mathrm{rad}} = \num{4.62e-5}(\frac{r}{\SI{1}{au}})^{-2}\,\si{.erg.cm^{-3}}$, and we can define the mean wavelength of radiation as $\bar{\lambdaup}=\int{\lambdaup u_\lambdaup d\lambdaup/u_{\mathrm{rad}}}=\SI{9.18e-5}{cm}$.

Inside the heliosphere $(r<\SI{100}{au})$, there are two main sources of hydrogen gas, one is from the solar wind and the other one is from the Local Interstellar Medium (LISM). The ROSAT mission provides a full sky coverage of 'local' X-ray emission of the LISM for the first time, and the properties of the LISM have been measured in the Ulysses mission \citep{Egger}. Inside the heliosphere, a few percent of the interstellar grain flux can be ionized by photo-ionization and charge-exchange with the solar wind \citep{geiss_origin_1996,Izmodenov2006}.

The number density of solar wind protons in the heliosphere can be approximated as $n_\mathrm{p}\mathrm{(SW)}=7(\frac{r}{\SI{1}{au}})^{-2}\,\si{.cm^{-3}}$ \citep{Gopalswamy2010,Baranov1995}, while the number density of hydrogen atoms from the LISM is approximately $n_\mathrm{H}\mathrm{(LISM)}=\SI{0.09}{cm^{-3}}$ \citep{Gopalswamy2010} for regions beyond $\SI{10}{au}$. Inside $\SI{10}{au}$ the contribution of hydrogen atoms from the LISM can be ignored compared to the solar wind proton number density. The temperature and drift velocity of solar wind protons and hydrogen atoms from the LISM are $T\mathrm{(SW)}\simeq\SI{8.3e4}{K}$, $T\mathrm{(LISM)}\simeq\SI{6306}{K}$ and $v\mathrm{(SW)}\simeq\SI{500}{km/s}$, $v\mathrm{(LISM)}\simeq\SI{26.24}{km/s}$, respectively \citep{Baranov1995,Izmodenov2006}. Since the mass of a proton and a hydrogen atom are approximately the same, the damping time of a dust grain in proton gas and hydrogen gas is also approximately the same (see Equation \ref{tH}). Therefore, it is convenient to use the weighted mean of these physical quantities:
\begin{equation}
    \bar{q}=\frac{n_\mathrm{p}\mathrm{(SW)}}{n_\mathrm{H}}q\mathrm{(SW)}+\frac{n_\mathrm{H}\mathrm{(LISM)}}{n_\mathrm{H}}q\mathrm{(LISM)}, \label{weighted}
\end{equation}
where $n_\mathrm{H}$ denotes the total number density of protons in the solar wind and hydrogen atoms in the LISM, and the physical quantity $q$ can represent either velocity or temperature.

In addition to gas, cosmic dust in the Solar System can be distinguished by its astronomical location: ISD, IDP and circumplanetary dust. However, in this paper, we only discuss ISD and IDP, as the trajectory of circumplanetary dust is rather complex. Moreover, the rotational damping time of circumplanetary dust depends on the atmosphere of the planets where it is located. ISD consists of dust in the Local Interstellar Cloud (LIC) that passes through the Solar System due to the relative velocity between the Solar System and the LIC. The Solar System is currently moving through the LIC at a velocity of $\sim\SI{26}{km.s^{-1}}$ approximately \citep{McComas2015}. At the boundary of the heliosphere, ISD particles with high charge-to-mass ratio are filtered out by the Lorentz force, and simulations have shown that ISD particles smaller than $\SI{0.01}{\mu m}$ are completely excluded \citep{Slavin2012,Sterken2019}. Regarding the origins of IDP, several studies suggest that they are brought from the asteroid belt, comets, and the Kuiper belt \citep{Mann2017,keller_evidence_2022}. Additionally, IDPs that are primarily located within about \SI{3}{au} from the Sun form zodiacal dust bands, which causes the zodiacal light \citep{jorgensen_distribution_2021}. 

The following equation represents the cumulative flux of measurement result at $r_0=\SI{1}{au}$ \citep{Mann2004}:
\begin{align} \label{flux_eq}
    F(m,r_0) = &(c_1m^{g_1}+c_2)^{g_2}+c_3(m+c_4m^{g_3}+c_5m^{g_4})^{g_5}\\ \nonumber
    &+c_6(m+c_7m^{g_6})^{g_7},
\end{align}
with $c_1=\num{2.2e3}$, $c_2=15$, $c_3=\num{1.3e-9}$, $c_4=\num{e11}$, $c_5=\num{e27}$, $c_6=\num{1.3e-16}$, $c_7=\num{e6}$, $g_1=\num{0.306}$, $g_2=\num{-4.38}$, $g_3=\num{2}$, $g_4=\num{4}$, $g_5=\num{-0.36}$, $g_6=\num{2}$, $g_7=\num{-0.85}$, where $F(m,r_0)$ is given in units of $\si{m^{-2}.s^{-1}}$, and the unit of $m$ is gram.
Regarding the radius dependence of the number density of dust grains, it is possible that dust grains within the zodiacal cloud originate from distinct sources. The number density of dust grains changes according to different heliocentric distances \citep{rowan2013improved,jorgensen_distribution_2021}. For simplicity, in this paper, we assume $\nu=1.3$ for all particle sizes over the entire heliosphere, with the number density proportional to $r^{-\nu}$ \citep{divine_five_1993,leinert_zodiacal_1981}. We further simplify the problem by considering only dust grains on orbits with low eccentricity, regardless of whether it is IDP or ISD. Consequently, the flux variation for low-eccentricity orbits can be approximated by extrapolation:
\begin{equation} \label{flux_extra}
    F(m,r) = F(m,r_0)(r/r_0)^{-1.8}.
\end{equation}

\subsection{Theory of Radiative Torque}
The RAT arises from the interaction of an anisotropic radiation field with an irregular dust grain. For a dust grain of size $a$, measured by the radius of a sphere of equal volume, the RAT is \citep{Draine1996}:
\begin{equation} \label{eq:Tau}
    \Gamma_\mathrm{RAT} = \pi a^2\gamma_{\mathrm{rad}} u_{\mathrm{rad}} \left(\frac{\bar{\lambdaup}}{2\pi}\right)\left\langle Q_\Gamma\right\rangle,
\end{equation}
where $\bar{\lambdaup}=\int{\lambdaup u_\lambdaup d\lambdaup/u_{\mathrm{rad}}}$ and $\gamma_{\mathrm{rad}}$ is anisotropy degree, and $\left\langle.\right\rangle$ denotes spectral averaging:
\begin{equation}
    \left\langle Q_\Gamma\right\rangle = \frac{\int{Q_\Gamma\lambdaup u_\lambdaup d\lambdaup}}{\int{\lambdaup u_\lambdaup d\lambdaup}}.
\end{equation}
Inside the heliosphere, $\gamma_{\mathrm{rad}}=1$, and the spectral averaging RAT efficiency, $\left\langle Q_\Gamma\right\rangle$ can be approximated to \citep{Hoang2019}:
\begin{equation}
     \left\langle Q_\Gamma\right\rangle \simeq \begin{cases}
     2\left(\frac{\bar{\lambdaup}}{a}\right)^{-2.7}\quad &a\lesssim\bar{\lambdaup}/1.8\\
     0.4\quad &a > \bar{\lambdaup}/1.8
     \end{cases} \label{RAT_eff}.
\end{equation}
Hence, the RAT, $\Gamma_\mathrm{RAT}$, becomes:
\begin{equation} \label{tauR}
     \Gamma_\mathrm{RAT} \simeq \begin{cases}
     \num{5.8e-29}a^{4.7}_{-5}\gamma_{\mathrm{rad}}U\bar{\lambdaup}_{0.5}^{-1.7}\,\si{.dyne.cm}\quad &a\lesssim\bar{\lambdaup}/1.8\\
     \num{8.6e-28}a^{2}_{-5}\gamma_{\mathrm{rad}}U\bar{\lambdaup}_{0.5}\,\si{.dyne.cm}\quad &a > \bar{\lambdaup}/1.8
     \end{cases},
\end{equation}
where $a_{-5}=a/\num{e-5}\,\si{.cm}$, $\bar{\lambdaup}_{0.5}=\bar{\lambdaup}/\SI{0.5}{\mu m}$.

\subsection{Rotational Damping Mechanisms and Timescales}
Rotational damping can be divided into several different processes, including neutral impacts, ion impacts, thermal emission of infrared photons, and rotational dipole emission \citep{DraineLazarian1998}.

Here, we calculate the general damping timescale of a dust grain damped by hydrogen gas as follows (see Appendix \ref{Appendix A}):
\begin{equation}
    \overline{\tau}_\mathrm{H}(v_\mathrm{d},v_\mathrm{th}) = \frac{3I}{2 n_\mathrm{H} m_\mathrm{H} a^4\overline{v_\mathrm{th}}\overline{K(s)}}, \label{tH}
\end{equation} 
where the bar notation denotes the weighted mean in Equation \ref{weighted} and will be omitted afterward, $n_\mathrm{H}$ is the total number density of protons, $v_\mathrm{d}$ is the drift velocity of dust grains with respect to the gas, $v_\mathrm{th}=(2kT_{\mathrm{gas}}/m_\mathrm{H})^{1/2}$ is the thermal gas velocity, $K(s) = 2\sqrt{\pi}+\pi s+\pi (e^{-s}-1)$ and $s = v_\mathrm{d}/v_\mathrm{th}$.

The above approximation is convenient for numerical calculation, as it covers both high and low sound speeds in one equation. For instance, for zodiacal dust, the average drift velocity is much larger than the thermal velocity for protons in the solar wind, $v_\mathrm{d} \gg \bar{v}_\mathrm{th}$ for $T_{\mathrm{gas}}=\SI{83000}{K}$ and $v_\mathrm{d}=\SI{500}{km/s}$. On the other hand, the average drift velocity and thermal velocity are around \SI{26}{km/s} according to Lyman-$\alpha$ observations \citep{Izmodenov2006} for dust colliding with neutral gas in the LISM which dominates within $\SI{10}{au}$ to $\SI{100}{au}$. Therefore, this example shows the importance of considering dust grains colliding with different sources of gas.

To facilitate comparisons across different sources of rotational damping, we adopt $\tau_\mathrm{H}$ with $v_\mathrm{d}=0$ as a fiducial timescale. The damping time due to process $j$ is expressed as:
\begin{equation}
    \tau_j^{-1}=F_j\tau_\mathrm{H}^{-1},
\end{equation}
and the total damping time is:
\begin{equation}
    \tau_\mathrm{damp}^{-1}=\left (F_n+F_i+F_\mathrm{IR}+F_\mathrm{ed} \right )\tau_\mathrm{H}^{-1},
\end{equation}
where $F_n$, $F_i$, $F_\mathrm{IR}$, and $F_\mathrm{ed}$ are the contributions from neutral imparts, ion impacts, thermal emission of infrared photons, and rotational dipole emission, respectively. For gas with He of 10\% abundance, $F_n+F_i\simeq 1.2$ \citep{DraineLazarian1998}. 

The contributions due to the rotational dipole emission ($F_\mathrm{ed}$) can be rewritten as \citep{DraineLazarian1998,hoang_spinning_2016}:
\begin{equation} \label{Fed}
    F_\mathrm{ed} = \left (\frac{I\omega^2}{3kT_\mathrm{d}}\right)\frac{\tau_\mathrm{H}}{\tau_\mathrm{ed}},
\end{equation}
where $T_\mathrm{d}$ is the dust temperature, and characteristic damping time $\tau_\mathrm{ed}$ is defined as:
\begin{equation} \label{ted}
    \tau_\mathrm{ed}=\frac{3 I^2c^3}{6 kT_\mathrm{gas}\mu_e^2}.
\end{equation}
As we will see in Section \ref{Rotational disruption of dust grains}, a dust grain is disrupted when its angular velocity, $\omega$, exceeds the critical rotational velocity, $\omega_\mathrm{disr}$. To estimate the upper bound of the rotational dipole emission, $F_\mathrm{ed}$, we substitute $\omega$ with $\omega_\mathrm{disr}$. By incorporating Equations \ref{tH}, \ref{ted}, and \ref{omega_disr} into Equation \ref{Fed}, the expression for $F_\mathrm{ed}$ is derived as:
\begin{align} \label{Fed2}
    F_\mathrm{ed} &= \num{1.4e-3} \left(\frac{S_{\mathrm{max},9}}{a_{-5}^{2}\hat{\rho}}\right)\left(\frac{\SI{30}{cm^{-3}}}{n_\mathrm{H}}\right)\left(\frac{T_\mathrm{d}}{\SI{15}{K}}\right)^{-1}\left(\frac{V}{\SI{3}{V}}\right)^2 \nonumber \\
    &\times\left(\frac{T_\mathrm{gas}}{\SI{100}{K}}\right)^{1/2}\left(\frac{2\sqrt{\pi}}{K(s)}\right),
\end{align}
where $a_{-5}=a/(\SI{e-5}{cm})$, $\hat{\rho}=\rho/(\SI{3}{g.cm^{-3}})$, and $S_{\mathrm{max},9}=S_{\mathrm{max}}/(\SI{e9}{erg.cm^{-3}})$.

Finally, the IR damping coefficient is \citep{DraineLazarian1998}:
\begin{equation}
    F_\mathrm{IR}\approx\left(\frac{0.4 U^{2/3}}{a_{-5}}\right)\left(\frac{\SI{30}{cm^{-3}}}{n_\mathrm{H}}\right)\left(\frac{\SI{100}{K}}{T_{\mathrm{gas}}}\right)^{1/2}\left(\frac{2\sqrt{\pi}}{K(s)}\right). \label{FIR}
\end{equation}
Notably, $\tau_\mathrm{IR}=F_\mathrm{IR}\tau_\mathrm{H}^{-1}$ remains independent of gas temperature $T_{\mathrm{gas}}$ and density $n_\mathrm{H}$. Furthermore, as demonstrated in \citet{Hoang2019} and \citet{Lazarian2007a}, for a dust grain that is trapped in high-$J$ attractor points, equilibrium rotation is attained at:
\begin{equation}
    \omega_\mathrm{RAT}=\frac{\Gamma_\mathrm{RAT}\tau_\mathrm{damp}}{I},
\end{equation}
where $I=8\pi\rho a^5/15$, $\rho=\SI{3}{g.cm^{-3}}$ are the moment of inertia and the typical density of the grain, respectively.

\begin{figure}
\includegraphics[width=\columnwidth]{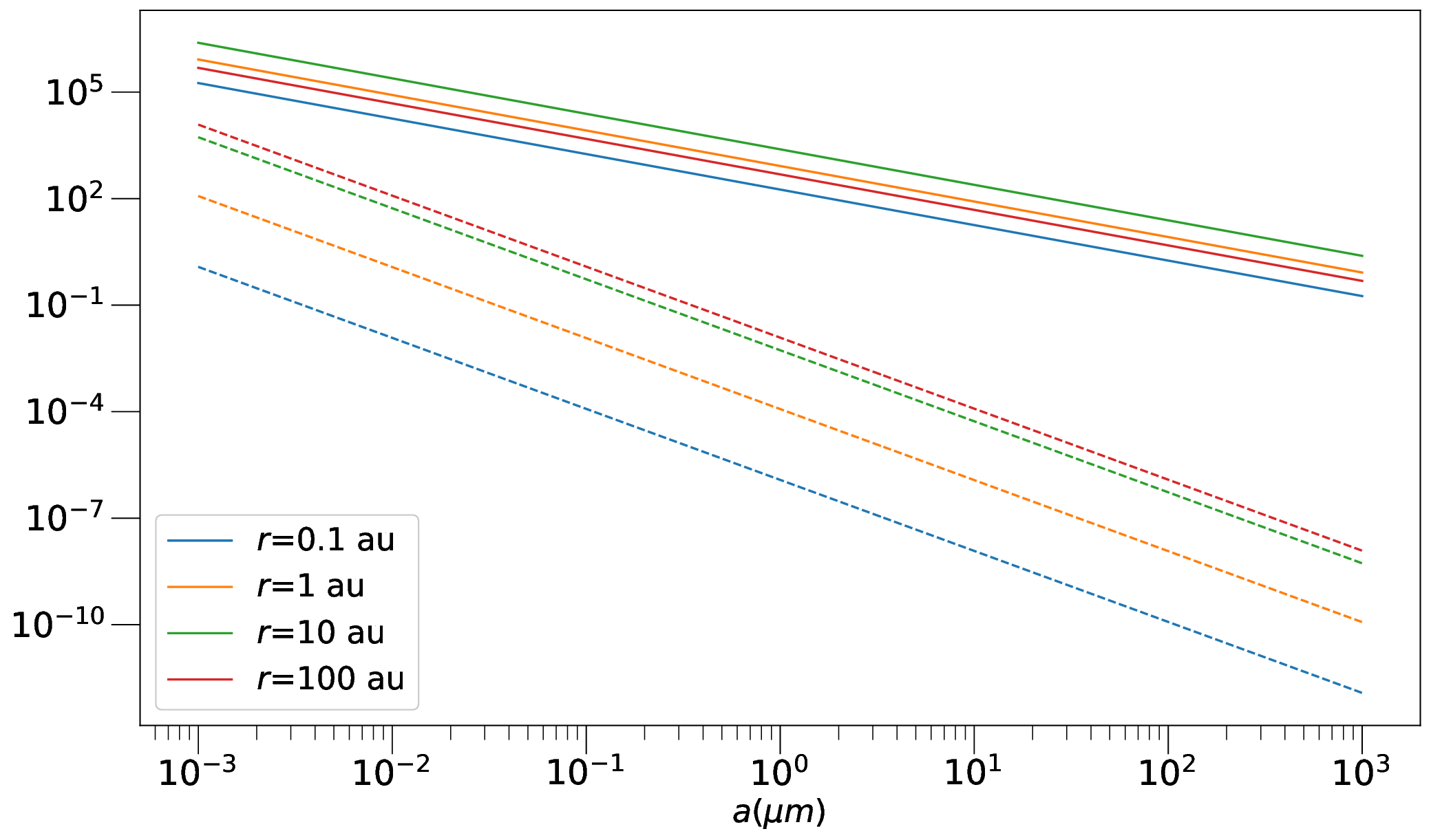}
\caption{Comparison of damping coefficient  due to thermal emission of infrared photons ($F_\mathrm{IR}$, solid lines; see Equation \ref{FIR}) and rotational dipole emission ($F_\mathrm{ed}$, dashed lines; see Equation \ref{Fed2}) for different grain sizes ($a$) and heliocentric distances ($r$).}
\centering \label{Taued}
\end{figure}

Using the parameters $S_{\mathrm{max},9} = 1$, $\hat{\rho} = 1$, and $T_\mathrm{d} = \SI{15}{K}$, Fig. \ref{Taued} compares the damping coefficient due to thermal emission of infrared photons ($F_\mathrm{IR}$, solid lines; see Equation \ref{FIR}) and rotational dipole emission ($F_\mathrm{ed}$, dashed lines; see Equation \ref{Fed2}) across different grain sizes ($a$) and heliocentric distances ($r$). The figure illustrates that, within heliocentric distances ranging from $\SI{0.1}{au}$ to $\SI{100}{au}$ and for grain sizes larger than $\SI{0.1}{\mu m}$, $F_\mathrm{IR}$ dominates over $F_\mathrm{ed}$. Consequently, $F_\mathrm{ed}$ can be neglected in the regions under consideration.

\subsection{Critical Rotational Velocity and Disruption Thresholds} \label{Rotational disruption of dust grains}
When a dust grain of size $a$ rotating at an angular velocity $\omega$, it is subjected to a centrifugal stress, defined as $S=\rho a^2\omega^2/4$ \citep{Hoang2019}. The critical rotational velocity occurs when this stress equals the maximum tensile strength $S_{\mathrm{max}}$, which depends on the composition and structure of the dust grain. This critical rotational velocity is given by:
\begin{align}
    \omega_\mathrm{disr}&=\frac{2}{a}\left(\frac{S_{\mathrm{max}}}{\rho}\right)^{1/2}\nonumber\\
    &\simeq\num{3.6e9}a_{-5}^{-1}\hat{\rho}^{-1/2}S_{\mathrm{max},9}^{1/2}\,\si{.rad/s} \label{omega_disr},
\end{align}

where $a_{-5}=a/(\SI{e-5}{cm})$, $\hat{\rho}=\rho/(\SI{3}{g.cm^{-3}})$, and $S_{\mathrm{max},9}=S_{\mathrm{max}}/(\SI{e9}{erg.cm^{-3}})$. The value of $\hat{\rho}$ and $S_{\mathrm{max}}$ can vary significantly among different dust grains compositions and structures. For reference, the density and tensile strength of bulk ice are about $\hat{\rho}=0.306$ and $S_{\mathrm{max}}=\SI{1e7}{dyn.cm^{-2}}$ \citep{Petrovic}, while cast iron exhibits $\hat{\rho}=2.4$ and $S_{\mathrm{max}}=\SI{2e9}{dyn.cm^{-2}}$ and monocrystalline silicon has $\hat{\rho}=0.776$ and $S_{\mathrm{max}}=\SI{7e10}{dyn.cm^{-2}}$. Additional values can be found in \cite{Howatson1972}. 

A dust grain will be disrupted by the centrifugal force if $\omega_\mathrm{disr}<\omega_\mathrm{RAT}$, that is:
\begin{equation}
    \frac{2}{a}\left(\frac{S_{\mathrm{max}}}{\rho}\right)^{1/2}<\frac{\Gamma_\mathrm{RAT}\tau_\mathrm{damp}}{I}.
\end{equation}
Thus, we have:
\begin{align}
    a_{-5}^{1.7}&>\num{7.02e-7}\hat{\rho}^{-1/2}S_{\mathrm{max},9}^{1/2}\gamma_{\mathrm{rad}}^{-1}\bar{\lambdaup}_{0.5}^{1.7}\times \label{a_rot_low} \\
    &\frac{n_\mathrm{H}(1.2+F_\mathrm{IR})v_\mathrm{th}K(s)}{U},\nonumber 
\end{align}
for grains with $a\lesssim\bar{\lambdaup}/1.8$, and
\begin{align}
    a_{-5}&<\num{2.1e7}\hat{\rho}^{1/2}S_{\mathrm{max},9}^{-1/2}\gamma_{\mathrm{rad}}\bar{\lambdaup}_{0.5}\times \label{a_rot_high} \\
    &\frac{U}{n_\mathrm{H}(1.2+F_\mathrm{IR})v_\mathrm{th}K(s)},\nonumber 
\end{align}
for grains with $a>\bar{\lambdaup}/1.8$.
Equation \ref{a_rot_low} and Equation \ref{a_rot_high} define the region in which the particles will be disrupted. The upper limit in Equation \ref{a_rot_high} arises from the larger moment of inertia associated with larger dust grains, while the lower limit in Equation \ref{a_rot_low} is caused by the smaller cross-sectional area of the dust grain.

In this study, we consider a strong solar radiation field within the heliosphere, i.e., $F_\mathrm{IR}\gg 1$. Equation \ref{a_rot_low} can be approximated as:
\begin{align}
    a_{\mathrm{disr,min}}\simeq\num{0.16}\left(\gamma_{\mathrm{rad}}^{-1}U^{-1/3}\bar{\lambdaup}_{0.5}^{1.7}S_{\mathrm{max},9}^{1/2}\right)^{1/2.7}\,\si{.\mu m}, \label{a_disr}
\end{align}
for grains with $a\lesssim\bar{\lambdaup}/1.8$. As expected, the minimum disruption grain size is independent of gas temperature and density in the first order approximation of Equation \ref{a_rot_low}.

Similarly, the maximum disruption grain size is found by solving Equation \ref{a_rot_high}:

\begin{align}
    a_{\mathrm{disr,max}}&\simeq \left(0.46\gamma_{\mathrm{rad}}\bar{\lambdaup}_{0.5}\hat{\rho}^{1/2}S_{\mathrm{max},9}^{-1/2}\frac{U}{\hat{n}_\mathrm{H}\hat{T}^{1/2} K(s)} \right.\\ 
    &\left.-\frac{U^{2/3}}{30\hat{n}_\mathrm{H}}\frac{2\sqrt{\pi}}{\hat{T}^{1/2}K(s)} \right) \,\si{.\mu m}, \nonumber
\end{align}
where $\hat{T}^{1/2}=\frac{T_{\mathrm{gas}}}{\SI{100}{K}}$ and $\hat{n}_\mathrm{H}=\frac{n_\mathrm{H}}{\SI{30}{cm^{-3}}}$. Under the strong radiation field assumption, the second term in RHS of the equation can be neglected, and thus we have:
\begin{align}
    a_{\mathrm{disr,max}}\simeq0.46\gamma_{\mathrm{rad}}\bar{\lambdaup}_{0.5}\hat{\rho}^{1/2}S_{\mathrm{max},9}^{-1/2}\frac{U}{\hat{n}_\mathrm{H}\hat{T}^{1/2}K(s)}\,\si{.\mu m}. \label{a_disr_max}
\end{align}
Notice that the maximum disruption grain size is dependent on gas temperature and density because the condition $F_\mathrm{IR} \gg 1$ is no longer valid for $a\approx a_{\mathrm{disr,max}}$. Furthermore, as detailed in \cite{Hoang2019_3}, the equation of motion for irregular grains under RATs and rotational damping is given by:
\begin{align} \label{EOM}
    \frac{Id\omega}{dt}&=\Gamma_\mathrm{RAT}-\frac{I\omega}{\tau_\mathrm{damp}}. 
\end{align}
Thus, the disruption time for rotational disruption with rotational damping is:
\begin{align} \label{t_disr}
    t_\mathrm{disr}&=-\tau_\mathrm{damp}\ln{\left(1-\frac{\omega_\mathrm{disr}}{\omega_\mathrm{RAT}}\right)}. 
\end{align}
We can define the characteristic timescale for rotational disruption in cases where the damping time greatly exceeds the disruption time ($\omega_\mathrm{RAT} \gg \omega_\mathrm{disr}$), or scenarios without rotational damping ($\tau_\mathrm{damp}\to\infty$). Hence, the characteristic timescale for rotational disruption can be estimated as:
\begin{align}
   t_{\mathrm{disr},0}&=\frac{I\omega_\mathrm{disr}}{\Gamma_\mathrm{RAT}}\\
    & \simeq \num{37.2}(\gamma U_7)^{-1}\bar{\lambdaup}_{0.5}^{1.7}\hat{\rho}^{1/2}S_{\mathrm{max},9}^{1/2}a_{-5}^{-0.7}\,\si{.days},  \label{tdisr01}
\end{align}
for $a\lesssim\bar{\lambdaup}/1.8$, and
\begin{align}
   t_{\mathrm{disr},0} \simeq \num{2.46}(\gamma U_7)^{-1}\bar{\lambdaup}_{0.5}^{-1}\hat{\rho}^{1/2}S_{\mathrm{max},9}^{1/2}a_{-5}^{2}\,\si{.days}, \label{tdisr02}
\end{align}
for $a>\bar{\lambdaup}/1.8$, where $U_7=U/\num{e7}$. It is worth noting that the criterion for rotational disruption, i.e. $\omega_\mathrm{disr}<\omega_\mathrm{RAT}$ implies that $t_{\mathrm{disr},0}<\tau_\mathrm{damp}$.

\subsection{Impact of Ice Mantles on Rotational Disruption} \label{Rotational disruption of ice mantles}
The ice mantle structure can form through heterogeneous condensation processes in low-temperature interstellar clouds \citep{Seki1983}. Detailed studies of these icy layer structures have been conducted by \cite{Hoang2020}. A layer structure forms when water vapor condenses on the surface of the grain core. For simplicity, a spherical dust grain of size $a$ is assumed. The average tensile strength of a rotating sphere measured at a distance $R$ from the centre of the dust grain is given by \citep{Kadish2005}:
\begin{align}
    S(R)=\frac{\rho\omega^2(a^2-R^2)}{4}.
\end{align}
Here, we assume the adhesion strength equals the tensile strength of ice, with the tensile strength of the core significantly higher than that of the outer layer. Consequently, the outer layer will break first due to centrifugal force, and the critical rotational velocity is given by $S=S_{\mathrm{max}}$.
\begin{align}
    \omega_\mathrm{disr}&=\left(\frac{4S_{\mathrm{max}}}{\rho(a^2-R^2)}\right)^{1/2}\nonumber\\
    &\simeq\num{3.6e9}(a^2-R^2)^{-1/2}a_{-5}^{-1}\hat{\rho}^{-1/2}S_{\mathrm{max},9}^{1/2}\,\si{.rad/s}.
\end{align}
Compared to Equation \ref{omega_disr}, we find that the tensile strength of ice mantles modifies to $S_{\mathrm{eff}}=S_{\mathrm{max}}/(1-(R/a)^2)$. Substituting $S_{\mathrm{max}}$ for $S_{\mathrm{eff}}$ in Equation \ref{a_disr} and Equation \ref{a_disr_max}, we derive the maximum and minimum disruption sizes in the ice mantle model:
\begin{align}
    a_{\mathrm{disr,min}}\simeq&\frac{\num{0.16}}{\left[1-(R/a)^2\right]^{1/5.4}}\times \nonumber\\ &\left(\gamma_{\mathrm{rad}}^{-1}U^{-1/3}\bar{\lambdaup}_{0.5}^{1.7}S_{\mathrm{max},9}^{1/2}\right)^{1/2.7}\,\si{.\mu m}, \label{a_disr_man}
\end{align}
\begin{align}
    a_{\mathrm{disr,max}}\simeq&0.46\gamma_{\mathrm{rad}}^{-1}\bar{\lambdaup}_{0.5}\hat{\rho}^{1/2}\left[1-(R/a)^2\right]^{1/2}S_{\mathrm{max},9}^{-1/2}\times \nonumber\\
    &\frac{U}{\hat{n}_\mathrm{H}}\frac{2\sqrt{\pi}}{\hat{T}^{1/2}K(s)}\,\si{.\mu m}.
    \label{a_disr_max_man}
\end{align}
Consequently, the minimum disruption size of a grain in the ice mantle model is higher than in the bulk ice model, while the maximum disruption size is lower. Since we can derive the result of ice mantles grain from the result of ice sphere grain by modifying the tensile strength $S$, we calculate the numerical result with ice sphere grain for simplicity, i.e., $R=0$.

\subsection{Poynting-Robertson Drag and Its Effect} \label{PR drag}
The Poything-Robertson (PR) drag can effectively remove nanoparticles, and we calculate the timescale of PR drag for comparison. The PR drag arises from momentum transfer due to solar wind particles and radiation pressure force, termed as plasma PR drag and photon PR drag, respectively. By incorporating the flux differential factor, which accounts for the difference between incoming and outgoing fluxes \citep{Grun1985}, into the PR drag timescale \citep{hoang_effect_2021}, the net mass loss timescale for a dust grain of size $a$ in a circular debris disk subject to the photon differential PR drag at distance $r$ from the Sun is given by:
\begin{equation} \label{t_PR}
    t_{\mathrm{ph}}=\frac{4\pi\rho r^2ac^2}{3\left\langle Q_{\mathrm{pr}}\right\rangle L_\odot}\frac{1}{\nu-1}\approx 697\frac{a_{-5}}{\left\langle Q_{\mathrm{pr}}\right\rangle}\left(\frac{r}{\SI{1}{au}}\right)^2\,\si{yrs}.
\end{equation}

Here, $L_\odot$ denotes the total solar energy emitted per second, $\nu=1.3$ is the exponent of the number density of dust grains as detailed in Section \ref{helio_Properties}. The parameters $\rho$ and $a$ represent the density and size of the dust grain, respectively, and $c$ is the speed of light. The average pressure cross-section efficiency $\left\langle Q_{\mathrm{pr}}\right\rangle$ depends on the shape and the composition of the dust grain \citep{zubko_light_2013}. Theoretical values of $\left\langle Q_{\mathrm{pr}}\right\rangle$ range from 0 for perfect transmitters to 2 for perfect backscatters, with unity for an ideal absorber \citep{krivov2006dust}. For particles with a small size parameter $x=2\pi a/\lambdaup \ll 1$, the pressure cross-section efficiency is approximately given by (See Appendix \ref{Appendix B}):
\begin{align}
    Q_{\mathrm{pr}}\simeq &-4\left ( \frac{2\pi a}{\lambdaup}\right )\Im{\frac{m^2-1}{m^2+2}},
\end{align}
where $m$ is the refractive index of the particle.

On the other hand, the ratio of the net mass loss timescale due to plasma differential PR drag $t_{\mathrm{sw}}$ to that of photon differential PR drag $t_{\mathrm{ph}}$ for a dust grain of size $a$ initially in a circular debris disk is given by \citep{minato_momentum_2006}:
\begin{equation}
    \frac{t_{\mathrm{sw}}}{t_{\mathrm{ph}}}=3\left(\frac{\dot{M}}{\dot{M}_\odot}\right)\left(\frac{L_\odot}{L}\right)\left(\frac{Q_{\mathrm{pr}}}{Q_{\mathrm{sw}}}\right),
\end{equation}
where $\dot{M}$ is the mass loss rate of the star and $\dot{M}_\odot$ is the mass loss rate of the Sun.

Consequently, the net mass loss timescale due to plasma differential PR drag in the heliosphere can be calculated as:
\begin{equation}
    t_{\mathrm{sw}}=2092\frac{a_{-5}}{\left\langle Q_{\mathrm{sw}}\right\rangle}\left(\frac{r}{\SI{1}{au}}\right)^2\,\si{yrs}.
\end{equation}

And the momentum transfer cross-section efficiency $Q_{\mathrm{sw}}$ is given by \citep{minato_momentum_2004}:
\begin{equation}
     Q_{\mathrm{sw}} = \begin{cases}
     \frac{2}{3}X \quad &X \leq 1\\
     1-\frac{1}{3X^2} \quad &X > 1
     \end{cases},
\end{equation}
where $X=2a/l(p_0)$ is the "size parameter"\footnote{Be careful not to confuse $X=2a/l(p_0)$ in plasma differential PR drag with $x=2\pi a/\lambdaup$ in photon differential PR drag.} and $p_0$ is the initial momentum of an incident ion. According to SRIM (Stopping and Range of Ions in Matter) calculations, the projected ranges of protons with initial kinetic energy $\SI{1}{keV/amu}$ for cases of \ch{SiO2} and ice are $\SI{0.07}{\mu m}$ and $\SI{0.11}{\mu m}$, respectively \citep{ziegler_srim_2010}. For simplicity, we adopt the projected range $l(p_0) = \SI{0.1}{\mu m}$.

Compared the timescale of plasma PR drag and photon PR drag, photon PR drag is more efficient for larger particles ($a>\SI{0.1}{\mu m}$), while plasma PR drag dominates for smaller particles ($a<\SI{0.1}{\mu m}$) \citep{minato_momentum_2004,Mann2007}. Consequently, the overall PR drag timescale is the minimum of $t_{\mathrm{ph}}$ and $t_{\mathrm{sw}}$ (e.g. $t_{\mathrm{PR}}=\min({t_{\mathrm{ph}},t_{\mathrm{sw}}})$), expressed as:
\begin{equation}
    t_{\mathrm{PR}}=2092\frac{a_{-5}}{Q_{\mathrm{eff}}}\left(\frac{r}{\SI{1}{au}}\right)^2\,\si{yrs}.
\end{equation}

where $Q_{\mathrm{eff}}=\max({3\left\langle Q_{\mathrm{pr}}\right\rangle,Q_{\mathrm{sw}}})$ is the effective cross-section efficiency.

\begin{figure}
\includegraphics[width=\columnwidth]{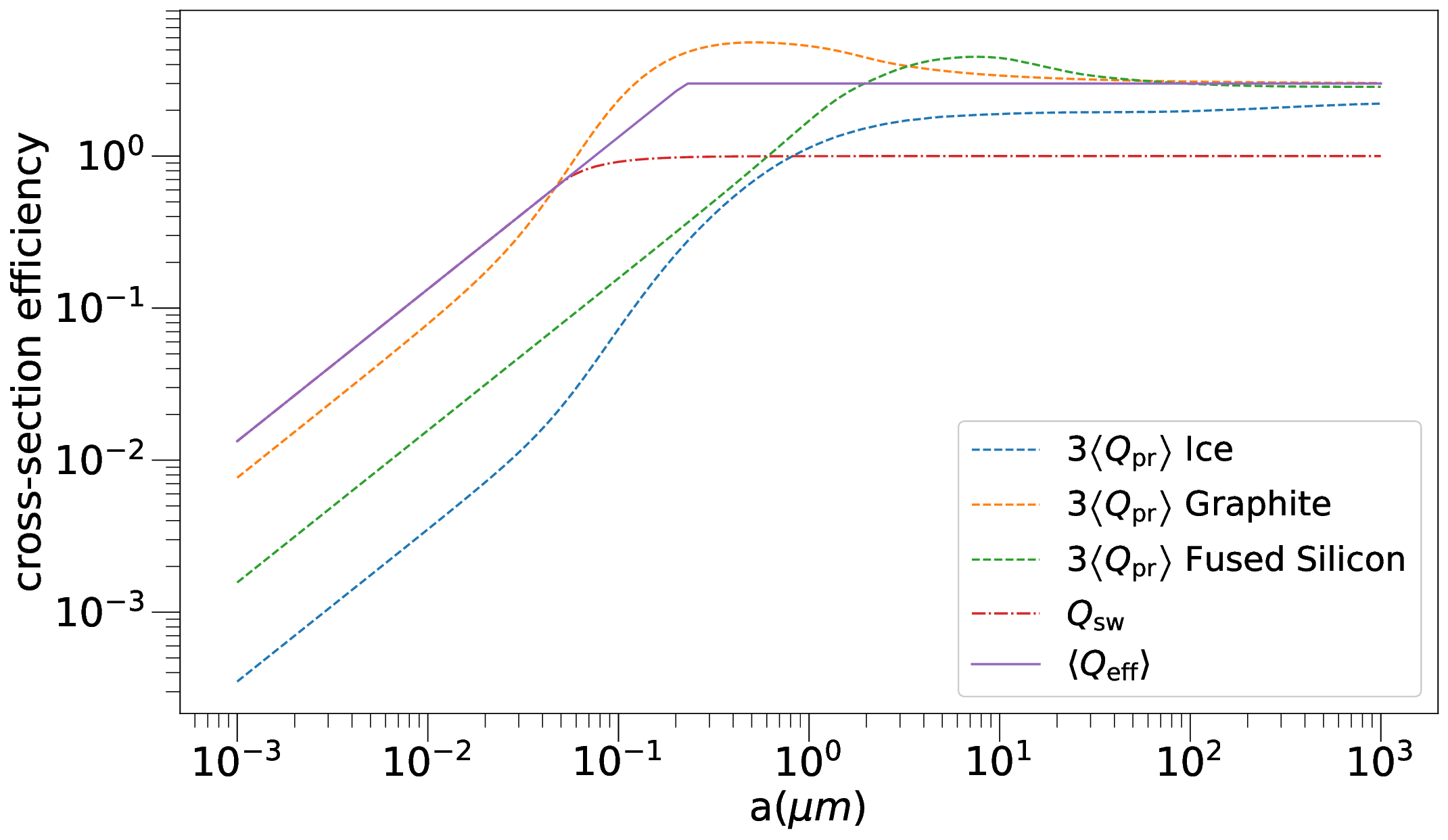}
\caption{Comparison of average pressure cross-section efficiency ($\left\langle Q_{\mathrm{pr}}\right\rangle$) of ice, graphite, and fused silicon (dashed line), momentum transfer cross-section efficiency ($\left\langle Q_{\mathrm{SW}}\right\rangle$) (dashed dot line), and the approximation value of the cross-section efficiency ($\left\langle Q_{\mathrm{eff}}\right\rangle$) (solid line). Pressure cross-section efficiencies are computed using optical constants of ice \citep{warren_optical_2008}, graphite \citep{djurisic_optical_1999}, and fused silicon \citep{kitamura_optical_2007}.}
\centering \label{cross_section}
\end{figure}

For convenient in numerical calculations, we approximate $Q_{\mathrm{eff}}$ as follows:
\begin{equation} \label{Qapp}
     Q_{\mathrm{eff}} = \begin{cases}
     \frac{2X}{3} \quad &X \leq \frac{9}{2}\\
     3 \quad &X > \frac{9}{2}
     \end{cases}.
\end{equation}
Fig. \ref{cross_section} compares the average pressure cross-section efficiency of ice, graphite and fused silicon (dashed line), the momentum transfer cross-section efficiency (dashed dot line), and the approximation value of $Q_{\mathrm{eff}}$ (solid line). The approximation closely matches $Q_{\mathrm{sw}}$ for small grains and $3\left\langle Q_{\mathrm{pr}}\right\rangle$ for large grains.

\subsection{Incorporating Rotational Disruption into Grain Size Distribution Models}
\label{sd_intro}
Once we obtain the timescales for different disruption and replenishment mechanisms, the size distribution can be computed using the model proposed by \citet{wyatt_debris_2011}. In their work, small size particles are replenished by the collisional disruption of larger particles, and the replenishment rate is determined by a redistribution function. Following their derivation, we include rotational disruption as an additional fragmentation mechanism and assume that the fraction of grains trapped in a high-$J$ attractor point is $f_{\mathrm{highJ}}=1$ \citep{herranen_alignment_2021} in this section. 

Consider a belt of particles where $m_k$ is the total mass in the $k$-th bin, $D_k$ is the typical size of the particles in that bin, and the bin width is defined by $D_{k+1}/D_{k}=1-\delta$. In this paper, we adopt a bin width of $\delta=0.002$, which satisfies the criterion $\delta << 1$ \citep{wyatt_debris_2011}. From mass conservation, we have:
\begin{equation}
    \dot{m}_k= \dot{m}^\mathrm{+c}_k - \dot{m}^\mathrm{-c}_k + \dot{m}^\mathrm{+r}_k - \dot{m}^\mathrm{-r}_k - \dot{m}^\mathrm{-pr}_k,
\end{equation}
where the subscript $k$ denotes the $k$-th bin, and the superscripts $+c$, $-c$, $+r$, $-r$, and $-pr$ refer to the mass gained from collisions in other bins, mass lost from collisions in these bins, mass gained from rotational disruption in other bins, mass lost from rotational disruption in these bins, and mass removed by PR drag, respectively.
We assume that the fragments follow a power-law distribution from sizes of $\eta_{\mathrm{rmax}} D_k$ to infinitesimally small particles with a power index $\alpha_r$, and that the redistribution function $F_r(k-i)$ represents the fraction of mass leaving the $i$-th bin from fragmentation that enters the $k$-th bin. This function is given by \citet{wyatt_debris_2011} as:
\begin{equation}
    F_r(l) = \eta^{\alpha_r-4}_{\mathrm{rmax}} (4-\alpha_r)\delta(1-\delta)^{l(4-\alpha_r)}.
\end{equation}
We assume that both collisional disruption and rotational disruption share the same cascade properties, i.e., $\eta_{\mathrm{rmax}}$ and $\alpha_r$. In this paper, we adopt $\alpha_r=3.5$ and $\eta_{\mathrm{rmax}}=2^{-1/3}$ \citep{wyatt_debris_2011}. Consequently, if we define the upper boundary as the bins where the typical particle size is $D_1$, and set the boundary condition such that the total loss rate from fragmentation disruption outside this boundary ($>D_1$) is constant, we can express this as:
\begin{equation} \label{loss_eq}
    \dot{m}^\mathrm{+c}_k + \dot{m}^\mathrm{+r}_k = \sum_{i=1}^{k-1}{(\dot{m}^\mathrm{-c}_i + \dot{m}^\mathrm{-r}_i)F_r(k-i)} + \dot{m}^{-}_\mathrm{b} \sum_{l=k}^{\infty}{F_r(l)}. 
\end{equation}
Here, the second term on the RHS of Equation \ref{loss_eq} indicates that the mass loss rate $\dot{m}^{-}_\mathrm{b}$ of the bins outside the upper boundary is constant. If we further assume that the size distribution is in a quasi-steady state, so that $\dot{m}_k=0$, then:
\begin{align} \label{loss_rate}
    \dot{m}^\mathrm{-c}_{ks} + \dot{m}^\mathrm{-r}_{ks} + \dot{m}^\mathrm{-pr}_{ks} =& \sum_{i=1}^{k-1}{(\dot{m}^\mathrm{-c}_{is} + \dot{m}^\mathrm{-r}_{is})F_r(k-i)} \\
    &+ \dot{m}^{-}_\mathrm{b} \sum_{l=k}^{\infty}{F_r(l)}, \nonumber
\end{align}
where the subscript $s$ denotes that the distribution is in steady state. We select $\dot{m}^{-}_\mathrm{b}=900 \delta\,\si{g/s} =\SI{1.8}{g/s}$ in an attempt to match the order of the number of the particles of the measurement result (Equation \ref{num_bin}) at the size $a=\SI{e2}{\mu m}$ with the situation where the loss rate due to rotational disruption is zero $R^r=0$. This choice of $\dot{m}^{-}_\mathrm{b}$ is motivated by the most massive population of dust of the size $\sim\SI{100}{\mu m}$ at $\SI{1}{au}$ \cite[e.g.,][]{Grun1985,love1993direct}. We note that despite this fitting to the measured the number of the particles at $\SI{e2}{\mu m}$, we do not intend to model all the measurement details, such as the impact energy of nanodust on wave antennas, as described in the Introduction. Our aim of this work is to explore potential influence of RATD on the dust evolution using a simple size distribution model.

The mass loss rate due to collision, $R_k^c=\dot{m}^\mathrm{-c}_k/m_k$, depends on the size distribution of dust grains and is given by \citet{wyatt_debris_2011}:
\begin{align} \label{Rk}
    R_k^c=\sum_{i=1}^{i_{ck}}{\frac{3m_i}{2\rho\pi D_i^3}(D_k+D_i)^2P_{ik}},
\end{align}
where $P_{ik}$ is the intrinsic collision probability between particles i and k, equal to $\pi v_\mathrm{rel}/V$. Here, $V=2 \pi^2 e^2 r^3$ is the volume through which the particles are moving, and $e=0.3$ is the average eccentricity of particles \citep{ipatov_dynamical_2008}. The index $i_{ck}$ denotes the bin with size $X_cD_k$, which corresponds to the size of the smallest particles causing catastrophic destruction. The parameter $X_c$ is given by:
\begin{align}
    X_c=(2Q^*_D/v_\mathrm{rel}^2)^{1/3},
\end{align}     
where $v_\mathrm{rel}\simeq e\sqrt{\frac{GM_\odot}{r}} \simeq 9\sqrt{\frac{\SI{1}{au}}{r}}\,\si{km/sec}$, $M_\odot$ is the mass of the Sun, and $Q^*_D$ is give by realistic dispersal laws \citep{wyatt_debris_2011}:
\begin{align}
    Q^*_D=Q_bD^{-b}+Q_cD^{c},
\end{align}
with $Q_b=\SI{790}{J.kg^{-1}}$, $b=0.38$, $Q_c=\SI{0.017}{J.kg^{-1}}$, $c=1.36$, and the unit of $D$ is meter. 

Furthermore, the mass loss rate for rotational disruption $R_k^r=\dot{m}^\mathrm{-r}_k/m_k$ and PR drag $R_k^\mathrm{-pr}=\dot{m}^\mathrm{-pr}_k/m_k$ are the inverse of $t_\mathrm{disr}$ in Equation \ref{t_disr} and $t_{\mathrm{PR}}$ in Equation \ref{t_PR}, respectively. The size distribution can be obtained by solving the following equation for $m_k$:
\begin{align} \label{mass_eq}
    m_k&=\frac{\dot{m}^\mathrm{-c}_k+\dot{m}^\mathrm{-r}_k+\dot{m}^\mathrm{-pr}_k}{R_k^c+R_k^r+R_k^\mathrm{-pr}}\nonumber\\
    &=\frac{\dot{m}^\mathrm{-c}_k+\dot{m}^\mathrm{-r}_k+\dot{m}^\mathrm{-pr}_k}{\tau_{\mathrm{c},k}^{-1}+t_{\mathrm{disr},k}^{-1}+t_{\mathrm{PR},k}^{-1}}.
\end{align}
This equation can be solved iteratively to find the steady state distribution. We compare the numerical results for the size distribution when both collisional and rotational disruption are present to the result when only collisional disruption is presented, and the result is shown in Section \ref{Siz_distribution_result}. 

As Equation \ref{mass_eq} indicates, the size distribution is determined by the loss rates of different disruption mechanisms. Therefore, different disruption models will result in different size distributions. Furthermore, this model does not include the radiation pressure, which will blow out micrometer-sized particles. The potential influence of the Lorentz force on sub-micron dust particles is also excluded from this simplified model.

\begin{figure*}
\includegraphics[width=0.9\textwidth]{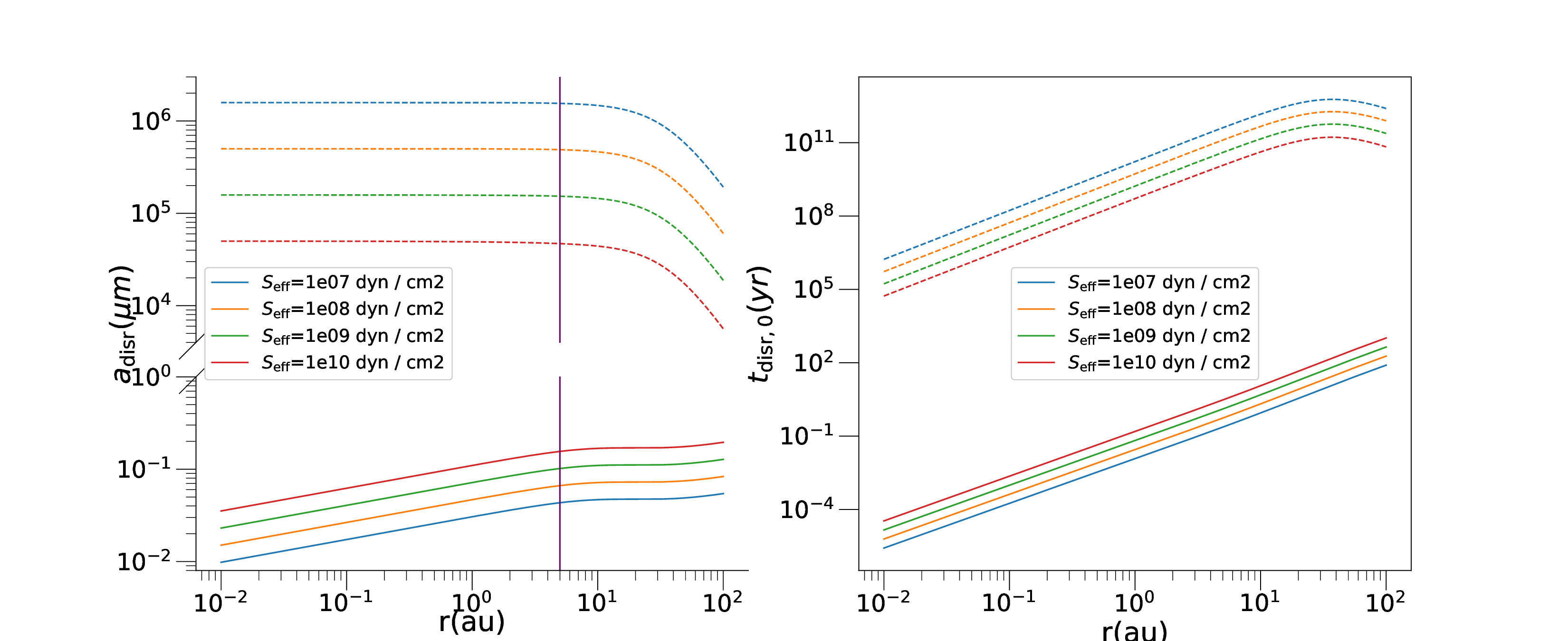}
\caption{The left panel shows the disruption grain size ($a_\mathrm{disr}$), as a function of the heliocentric distance ($r$) for the different effective tensile strengths ($S_{\mathrm{eff}}$), the upper and lower lines corresponding to Equation \ref{a_disr_max} and Equation \ref{a_disr}, respectively. The right panel shows the characteristic disruption time evaluated at $a=a_\mathrm{disr}$, the upper lines (dashed lines) and lower lines (solid lines) corresponding to Equation \ref{a_disr_max} and Equation \ref{a_disr}, respectively. The lower limit of the snow line is also plotted in the figure. The vertical purple line indicates the present snow line, which is defined by the temperature of the dust grains.}
\centering \label{disruption_size}
\end{figure*}

\subsection{Modeling the Distribution of Dust Grains in high-$J$ and low-$J$ Attractor Points}
\label{model_J}
In Section \ref{sd_intro}, we assumed that dust grains are always trapped at a high-$J$ attractor point. However, RATD cannot be so efficient. If grains subject to only radiative torques, i.e., disregarding magnetic torques, \citet{Lazarian2007a} showed that the parameter space of a grain can include scenarios where there are no high-$J$ attractor points, and \citet{herranen_alignment_2021} demonstrated that the fraction of grains trapped in a high-$J$ attractor point $f_{\mathrm{highJ}}$ is always lower than 1 in their ensemble of grain shapes. However, for grains with embedded iron inclusions, the value of $f_\mathrm{highJ}$ can reach $100\%$ due to the effect of enhanced magnetic relaxation \citep{hoang_unified_2016, herranen_alignment_2021}. Note that only grains at high-$J$ attractors can be disrupted by RATD. In Appendix \ref{Appendix C}, we estimate magnetic torques on heliospheric dust, and suggest that they could significantly vary $f_\mathrm{highJ}$ depending on dust size, charge distribution, and magnetic properties.

To keep things simple, unless otherwise stated, we consider $f_{\mathrm{highJ}}$ as a free parameter and adopt a fixed fraction $f_{\mathrm{highJ}}$ for all fragment sizes originating from the border and various fragmentation methods. Let $m_{\mathrm{H},k}$ and $m_{\mathrm{L},k}$ represent the mass of the dust grains trapped in a high-$J$ attractor point and a low-$J$ attractor point in the $k$-th bin, respectively. With reference to Equation \ref{loss_rate}, the following are the loss rate equations for $m_{\mathrm{H},k}$ and $m_{\mathrm{L},k}$: \\
For grains in a high-$J$ attractor point: 
\begin{align}
    \label{loss_rate_e1}
    \dot{m}^\mathrm{-c}_{\mathrm{H},ks} + \dot{m}^\mathrm{-r}_{\mathrm{H},ks} +& \dot{m}^\mathrm{-pr}_{\mathrm{H},ks} = \\
    &f_{\mathrm{highJ}}\sum_{i=1}^{k-1}{(\dot{m}^\mathrm{-c}_{is} + \dot{m}^\mathrm{-r}_{\mathrm{H},is})F_r(k-i)} \nonumber\\
    &+ f_{\mathrm{highJ}}\dot{m}^{-}_\mathrm{b} \sum_{l=k}^{\infty}{F_r(l)}. \nonumber
\end{align} \\
For grains in a low-$J$ attractor point: 
\begin{align} 
    \label{loss_rate_e2}
    \dot{m}^\mathrm{-c}_{\mathrm{L},ks} +& \dot{m}^\mathrm{-pr}_{\mathrm{L},ks} = \\
    &(1-f_{\mathrm{highJ}})\sum_{i=1}^{k-1}{(\dot{m}^\mathrm{-c}_{is} + \dot{m}^\mathrm{-r}_{\mathrm{H},is})F_r(k-i)}\nonumber\\
    &+ (1-f_{\mathrm{highJ}})\dot{m}^{-}_\mathrm{b} \sum_{l=k}^{\infty}{F_r(l)}. \nonumber
\end{align}

The size distribution of grains trapped in high-$J$ and low-$J$ attractor points can be obtained by solving the following equations for $m_{\mathrm{H},k}$ and $m_{\mathrm{L},k}$: \\
For grains in a high-$J$ attractor point: 
\begin{align} \label{mass_eq_e1}
    m_{\mathrm{H},k}=\frac{\dot{m}^\mathrm{-c}_{H,k}+\dot{m}^\mathrm{-r}_{H,k}+\dot{m}^\mathrm{-pr}_{H,k}}{\tau_{\mathrm{c},k}^{-1}+t_{\mathrm{disr},k}^{-1}+t_{\mathrm{PR},k}^{-1}}.
\end{align}\\
For grains in a low-$J$ attractor point: 
\begin{align} \label{mass_eq_e2}
    m_{\mathrm{L},k}=\frac{\dot{m}^\mathrm{-c}_{L,k}+\dot{m}^\mathrm{-pr}_{L,k}}{\tau_{\mathrm{c},k}^{-1}+t_{\mathrm{PR},k}^{-1}}.
\end{align}

And the total size distribution is the following:
\begin{align}
    m_{k}=m_{\mathrm{H},k}+m_{\mathrm{L},k}.
\end{align}
Therefore, the mass loss rate due to collisions $R_k^c$ is given by:
\begin{align} \label{Rk_2}
    R_k^c&=\sum_{i=1}^{i_{ck}}{\frac{3m_i}{2\rho\pi D_i^3}(D_k+D_i)^2P_{ik}} \\
    &= \sum_{i=1}^{i_{ck}}{\frac{3(m_{\mathrm{H},i}+m_{\mathrm{L},i})}{2\rho\pi D_i^3}(D_k+D_i)^2P_{ik}}. \nonumber
\end{align}

\section{Numerical Results}
\label{result}
In this section, we present the numerical results of the RATD theory applied to the Solar System. This includes the disruption sizes and characteristic timescales for various fragmentation mechanisms and different fractions of grain alignment. We also examine how the positions of the present water snow line shift with different grain sizes due to RATD effects in the Solar System. The computational domain for this analysis spans from $\SI{e-3}{\mu m}$ to $D_1 = \SI{e4}{\mu m}$, with the total mass of the bins outside the lower boundary extrapolated from the values within this range. Because the observed dust size distribution described in Equation \ref{flux_eq} was obtained at \SI{1}{au}, we focus on a heliocentric distance of \SI{1}{au} as the fiducial location to study the effect of RATD on the dust size distribution in Sections 3.2-3.3. Additionally, we will discuss the effect of radiation pressure, which is not included in our model, on dust grains affected by RAT in the heliosphere. 

\subsection{Disruption Sizes and Characteristic Times of Dust Grains}
\label{size and time}

\begin{figure*}
\includegraphics[width=0.9\textwidth]{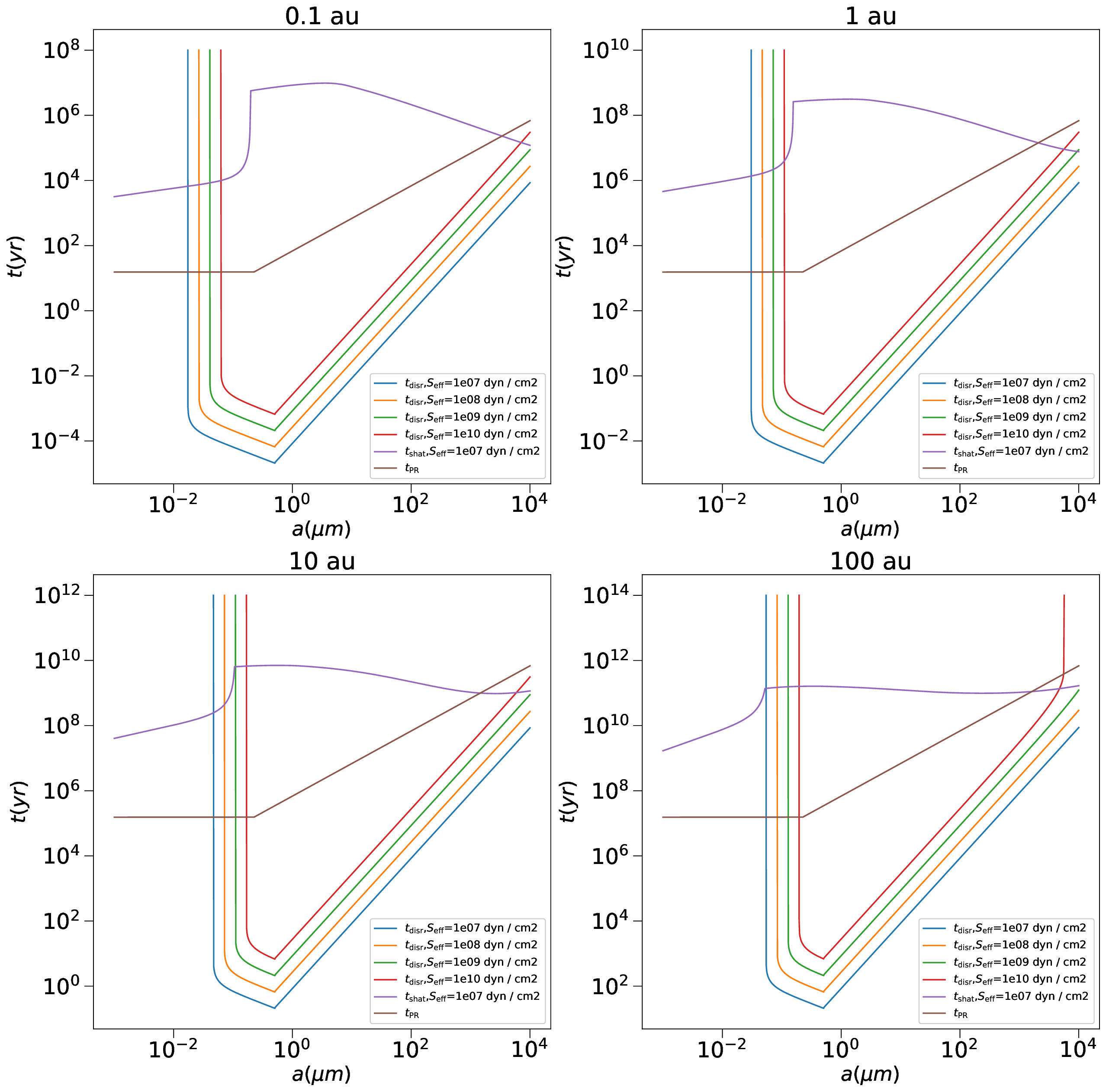}
\caption{The characteristic times of different mechanisms of fragmentation and tensile strengths, $S_\mathrm{eff}$, as a function of grain size at heliocentric distances of $\SI{0.1}{au}$, $\SI{1}{au}$, $\SI{10}{au}$ and $\SI{100}{au}$. The vertical curves arise because the timescale of rotational disruption diverges at $a_{\mathrm{disr,min}}$ and $a_{\mathrm{disr,max}}$. The plotted timescales include those for the rotational disruption ($t_\mathrm{disr}$), PR drag ($t_\mathrm{PR}$), and collisional disruption ($t_\mathrm{shat}$).}
\centering \label{characteristic_time}
\end{figure*}

Figure \ref{disruption_size} shows the minimum and maximum grain disruption sizes and their corresponding RATD timescales based on Section \ref{Rotational disruption of dust grains}. The left panel shows that the disruption size of dust grains ranges from about $\SI{0.01}{\mu m}$ to $\SI{0.1}{\mu m}$, and the disruption range is defined by the grain size between the minimum grain disruption sizes (solid line) and the maximum grain disruption sizes (dashed line). For $r>\SI{10}{au}$, the density of protons in the solar wind is smaller than those in the LISM, so the disruption size is mainly affected by the LISM. We also notice that the slope of the curves for the minimum and maximum disruption size $a_{\mathrm{disr,min}}$ changes slightly beyond $\SI{10}{au}$. The minimum disruption size changes because the drift velocity of protons in the LISM is lower than those in the solar wind, and therefore $K(s)$ is lower in Equation \ref{a_disr}, and the maximum disruption size is changed significantly because the proton density changes from $n_\mathrm{H}\propto r^{-2}$ to a constant (see Equation \ref{a_disr_max}). It is also interesting to see that the maximum disruption size can be up to $\SI{1}{m}$. However, it would take a very long time and will likely be destroyed by grain-grain collisions first (see Fig. \ref{characteristic_time}). For grain sizes smaller than $a_{\mathrm{disr,min}}$, since they cannot be disrupted by RATD, the PR drag will remove them on the timescale $t_{\mathrm{PR}}$.

In order to show the importance of rotational disruption, we compare the characteristic times of rotational disruption (Equation \ref{t_disr}), PR drag (Equation \ref{t_PR}), and collisional disruption (Equation \ref{Rk_2}) as a function of grain sizes at different heliocentric distances in Fig. \ref{characteristic_time}. The characteristic times of collisional disruption are calculated using the size distribution obtained in Section \ref{model_J} with $S_{\mathrm{eff}}=10^7 {\rm dyn cm^{-2}}$ and $f_{\mathrm{highJ}}=0.8$. The result shows the similarity of the overall trend for $r<\SI{100}{au}$. As expected, the timescale of rotational disruption diverges at $a_{\mathrm{disr,min}}$ and $a_{\mathrm{disr,max}}$, as indicated in Equation \ref{a_disr} and \ref{a_disr_max}. The timescale of rotational disruption is the shortest between size $a_{\mathrm{disr,min}}$ and $\SI{e4}{\mu m}$. Additionally, we can observe that all characteristic times increase as the heliocentric distance $r$ increases. Notably, there is a change of slope at dust size $a=\SI{0.51}{\mu m}$ for the timescale of rotational disruption, due to the reduction of RAT efficiency in Equation \ref{RAT_eff}. Furthermore, there is an abrupt change in the timescale of collisional disruption at the dust size $a$, where $X_c(a)a=a_{\mathrm{disr,min}}$. This abrupt change is due to the inclusion of the mass distribution variation in the summation in Equation \ref{Rk}.

It is worth mentioning that Fig. 5 shows the collisional timescale becomes larger than the PR timescale when the dust size $< \SI{e3}{\mu m}$. It is expected from our choice of the boundary condition ($\dot{m}^{-}_\mathrm{b}$) to roughly capture the most massive population in the dust of $\SI{100}{\mu m}$ at $\SI{1}{au}$. In other words, in the absence of RATD, the dust size evolution makes a transition from the collision dominated regime for large sizes ($> \sim \SI{100}{\mu m}$) to the PR drag dominated regime for small sizes ($< \sim \SI{100}{\mu m}$), in general agreement with theoretical interpretations \citep[e.g.,][]{Grun1985}.

\subsection{Effect of the Fraction of Grain Alignment on Grain Size Distribution}
\label{Grain_Alignment_result}
\begin{figure}
\includegraphics[width=\columnwidth]{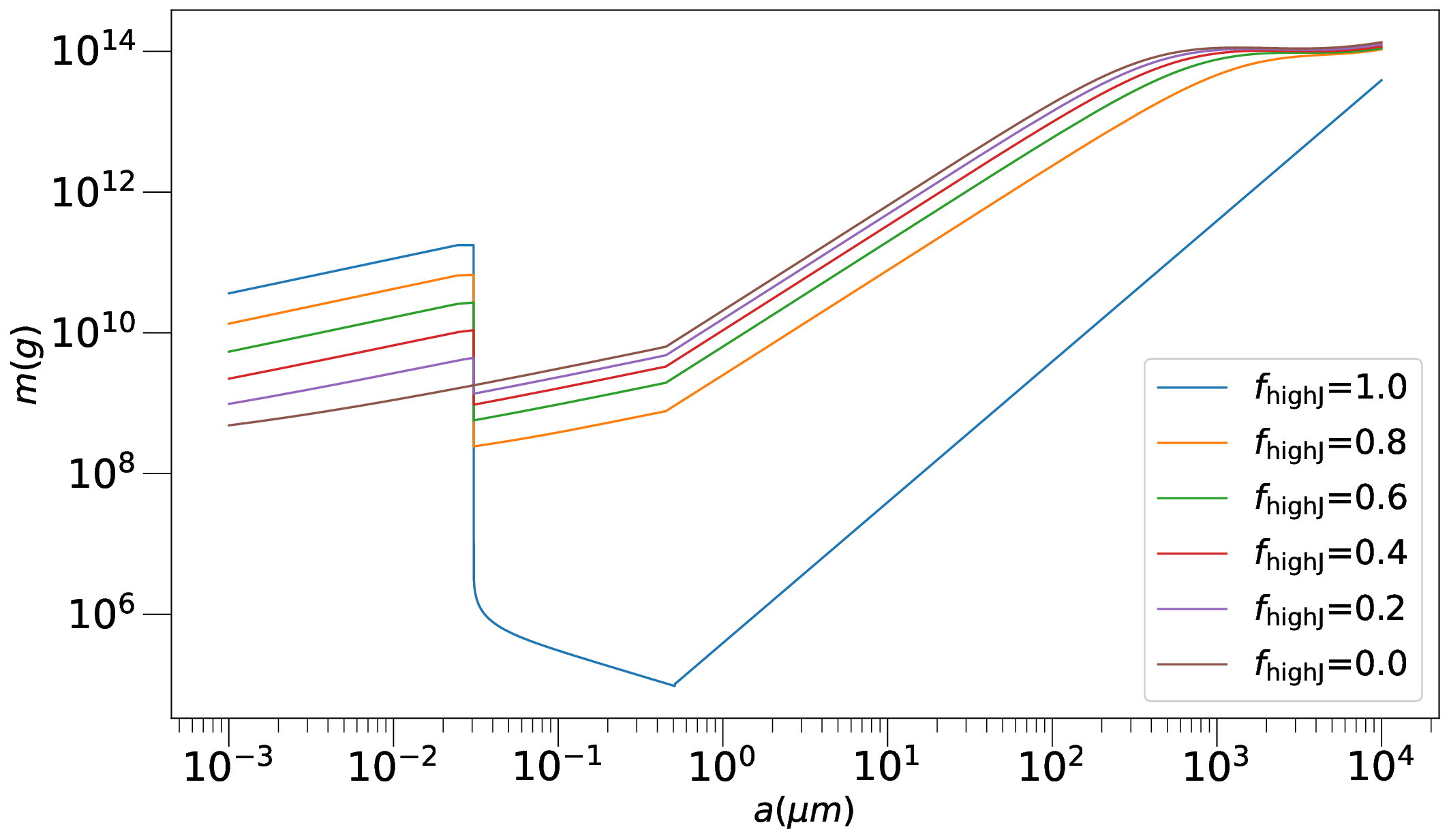}
\caption{Mass distribution of dust grains for different fractions of grain alignment ($f_{\mathrm{highJ}}$) and tensile strength ($S_{\mathrm{eff}}=10^7 {\rm dyn cm^{-2}}$). The mass distribution lies between the cases of $f_{\mathrm{highJ}}=0$ and $f_{\mathrm{highJ}}=1$ at $\fidloc$. The y-axis is the total mass ($m$) in logarithmic bins centered on sizes ($a$) given on the x-axis.}
\centering \label{size_distribution_fr}
\end{figure}

To account for the variation of the grain alignment at high-$J$ attractors with the grain size/charge/shapes and magnetic properties (\citealt{hoang_unified_2016,herranen_alignment_2021}, also see Appendix \ref{Appendix C}), we vary the fraction $f_{\rm highJ}$ from $0-1$. Fig. \ref{size_distribution_fr} shows the mass distribution for a range of $f_{\mathrm{highJ}}$ with a tensile strength of $S_{\mathrm{eff}}=10^7{\rm dyn cm^{-2}}$, and a mass loss rate of $\dot{m}^{-}_\mathrm{b}=\SI{1.8}{g/s}$ at $\fidloc$. It illustrates that the mass distribution falls between the cases of $f_{\mathrm{highJ}}=0$ and $f_{\mathrm{highJ}}=1$.

This figure shows that the influence of RATD on the size distribution of dust grains in the Solar System depends significantly on the fraction of grains trapped in high-$J$ attractor points. As $f_{\mathrm{highJ}}$ increases, the mass distribution for larger grain sizes ($a > a_{\mathrm{disr,min}}$) decreases, while the mass distribution for smaller grain sizes ($a < a_{\mathrm{disr,min}}$) increases. Moreover, despite the increasing fraction of aligned grains, the RATD-dominated regime maintains a mass distribution pattern closer to the scenario where no grains are trapped in high-$J$ attractor points ($f_{\mathrm{highJ}}=0$). As a result, the size distributions for $f_{\mathrm{highJ}}<0.8$ closely resemble those in the absence of rotational disruption, in contrast to the distributions observed when $f_{\mathrm{highJ}}=1$.

Fig. \ref{size_distribution_fr} also shows that there is saturation in the mass distribution within the size range of $\SI{e2}{\mu m}$ to $\SI{e3}{\mu m}$ all cases except when $f_{\mathrm{highJ}}=1$. This saturation results from a transition between a regime dominated by PR drag and a regime dominated by collisions for grains trapped in low-$J$ attractor points. In these regimes, the timescale changes from decreasing with size to increasing with size (see Fig. \ref{characteristic_time}).

\begin{figure}
\includegraphics[width=\columnwidth]{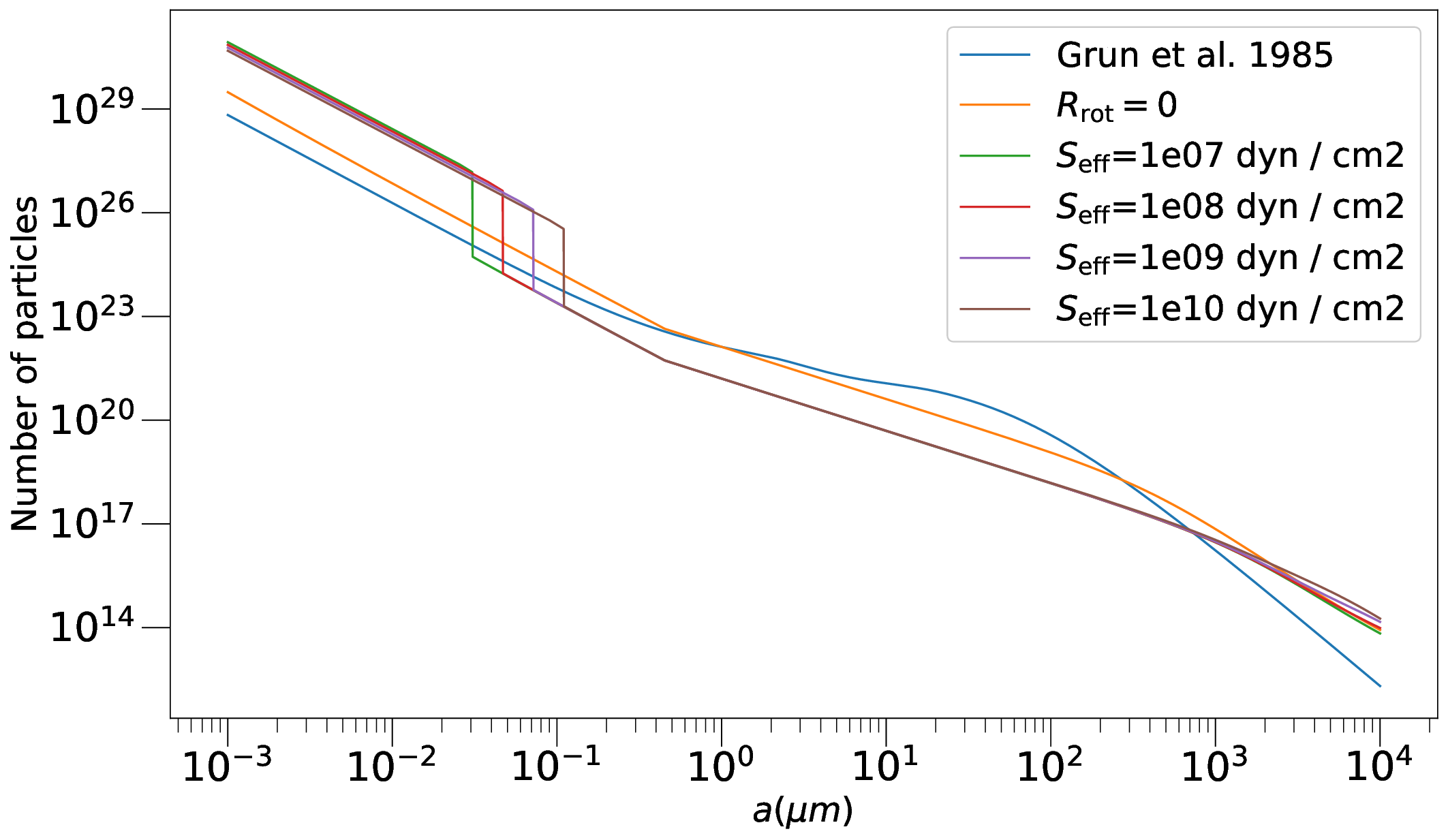}
\caption{The number of particles in each bins at $\fidloc$, with different tensile strengths ($S_{\mathrm{eff}}$) defined in Section \ref{sd_intro}. The size distribution without the rotational disruption, denoted by $R_\mathrm{rot} = 0$, is also plotted. The measurement curve at $\fidloc$ given by \citet{Grun1985} is also shown for comparison.}
\centering \label{Number}
\end{figure}

\begin{figure}
\includegraphics[width=\columnwidth]{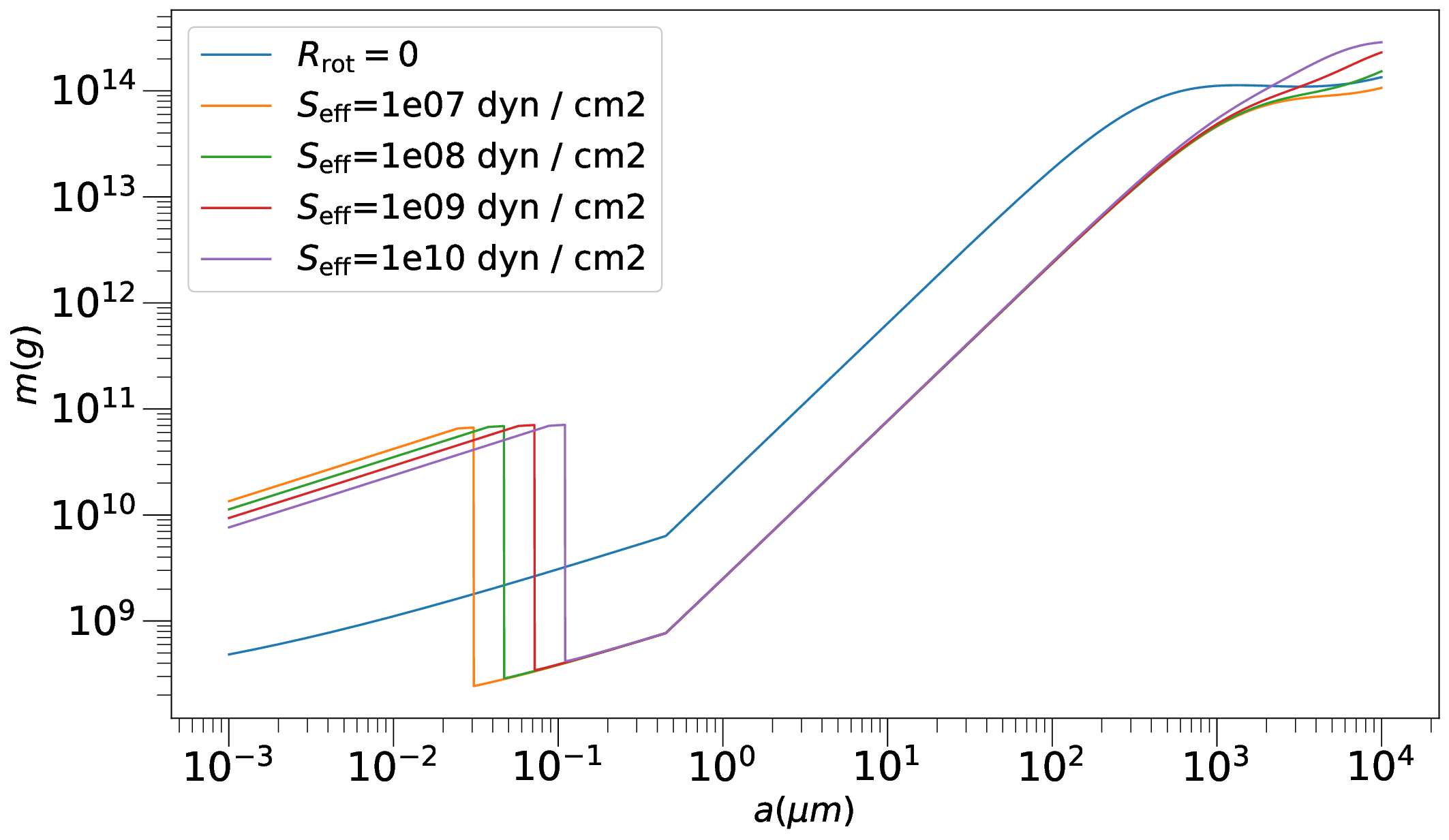}
\caption{The steady state grain size distribution for the different tensile strengths ($S_{\mathrm{eff}}$) with and without rotational disruption at $\fidloc$. The y-axis is the mass ($m$) in logarithmic bins centred on sizes ($a$) given on the x-axis. The case $R_\mathrm{rot} = 0$ corresponds to the scenario without rotational disruption.}
\centering \label{size_dist}
\end{figure}

\begin{figure}
\includegraphics[width=\columnwidth]{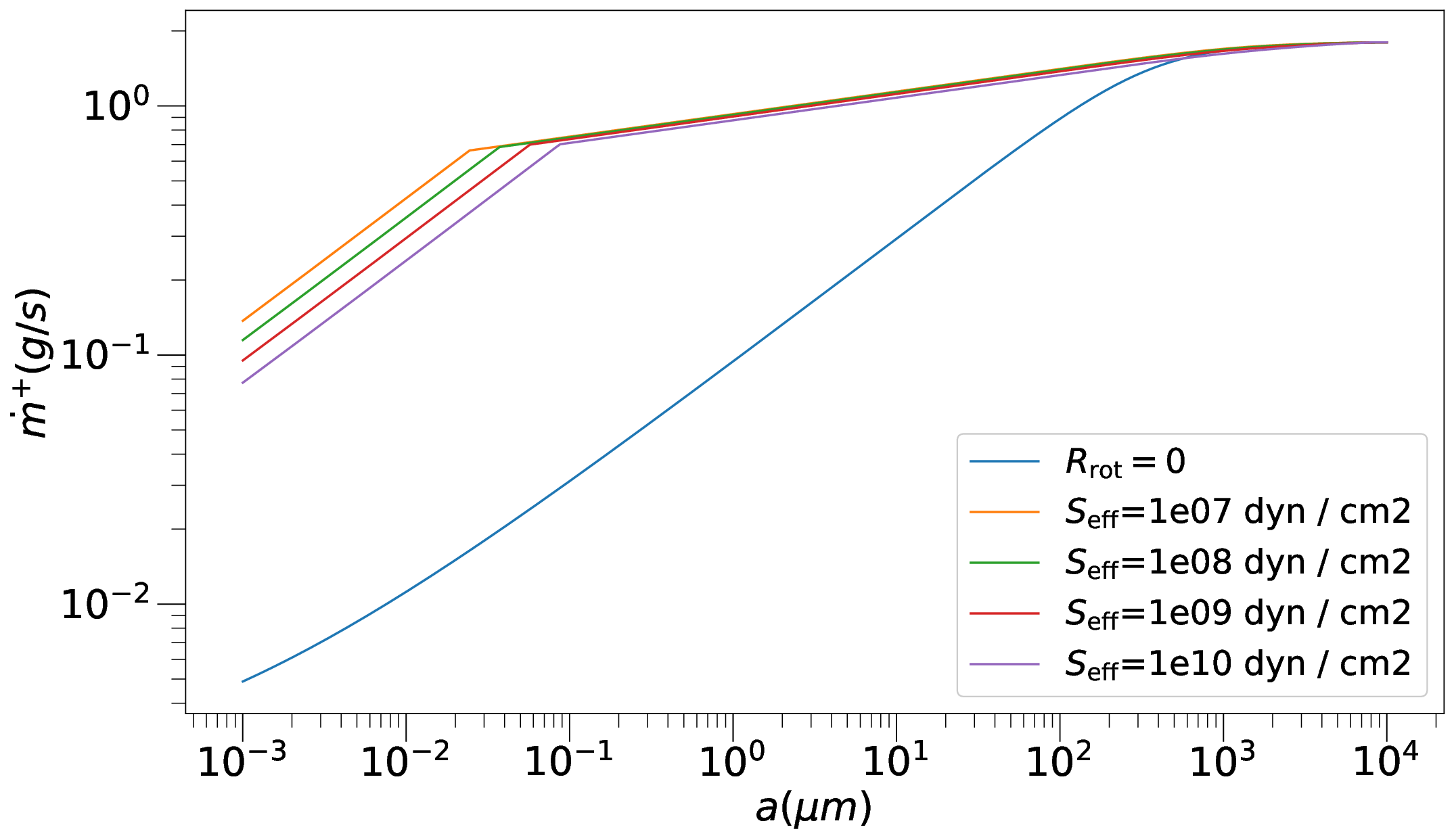}
\caption{The total mass gained versus different bin sizes with different tensile strengths ($S_{\mathrm{eff}}$) with and without rotational disruption at $\fidloc$  ($R_\mathrm{rot}=0$ for no rotational disruption). The y-axis is the total mass gained ($\dot{m}^{+}$) in logarithmic bins centred on sizes ($a$) given on the x-axis.}
\centering \label{gain_rate}
\end{figure}

\begin{figure}
\includegraphics[width=\columnwidth]{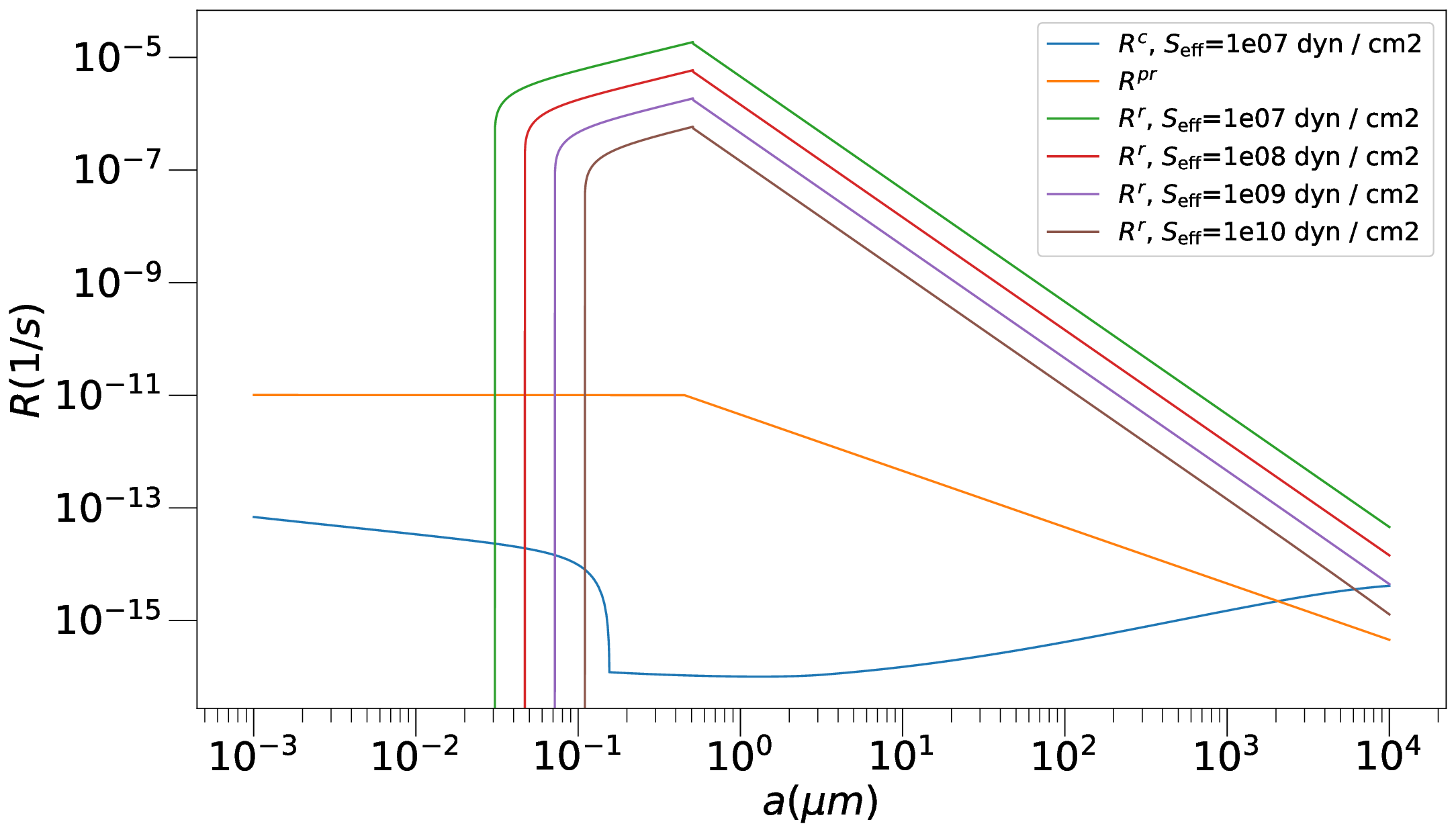}
\caption{The loss rate of collision ($R^c$), rotational disruption ($R^r$), and PR drag ($R^\mathrm{-pr}$) of different bins with different tensile strengths ($S_{\mathrm{eff}}$). The y-axis represents the loss rate ($R$) in logarithmic bins, with the bin centers corresponding to the sizes ($a$) indicated on the x-axis.}
\centering \label{lost_rate}
\end{figure}

\subsection{Effect of the Tensile Strength on Size Distribution} \label{Siz_distribution_result}
We now examine how different tensile strengths in RATD affect the steady-state size distribution. Following the methodology outlined in Section \ref{Grain_Alignment_result}, we use $D_1=\SI{e4}{\mu m}$ and $\dot{m}^{-}_\mathrm{b}=\SI{1.8}{g/s}$ for our fiducial location at $\fidloc$, with a high-$J$ fraction of $f_{\mathrm{highJ}}=0.8$. 

Based on Equation \ref{flux_eq}, the number of particles observed with size between $D_k$ and $D_{k+1}$ can be expressed as:
\begin{equation} \label{num_bin}
    N_k=-\frac{\partial F(m,r)}{\partial m}\frac{dm}{dD}\frac{V D_k \delta}{v_k}=-\frac{\partial F(m,r)}{\partial m}\frac{\pi}{2} D_k^3 \frac{V \rho \delta}{v_k},
\end{equation}
where $v_k = \sqrt{\frac{GM_\odot}{r}}$ represents the Keplerian velocity at the heliocentric distance $r$. Fig. \ref{Number} illustrates the number of particles in each bin based on numerical results, both with and without considering rotational disruption at $\fidloc$, along with the observational results derived from Equation \ref{num_bin} at $\SI{1}{au}$. The figure shows that the number of particles for larger dust particles ($> \SI{0.1}{\mu m}$) is more depleted by RATD than the case without RATD, thus dramatically increasing the number of particles for smaller dust particles ($< \SI{0.1}{\mu m}$) by rotational fragmentation. This abundance jump of sub-micrometer-sized particles occurs at a smaller $a$ for a weaker dust tensile strength $S_{\mathrm{eff}}$, as illustrated in Fig \ref{Number}. This trend is expected from Fig \ref{characteristic_time}.

Fig. \ref{size_dist} shows the steady-state size (or mass) distribution at $\fidloc$ with different tensile strengths, along with the steady-state size distribution without rotational disruption at $\fidloc$. The y-axis represents the mass in logarithmic bins centred on sizes given on the x-axis. As expected, the mass in the bins of micrometer size with rotational disruption is significantly lower than the size distribution without rotational disruption. 

To better understand why the mass in the bins of sub-micrometer size with rotational disruption is significantly higher than the size distribution without rotational disruption, we plot the total mass gain rate $\dot{m}^{+}_k = \dot{m}^\mathrm{+c}_k + \dot{m}^\mathrm{+r}_k$ in Fig. \ref{gain_rate} and the loss rate of different mechanisms and tensile strengths $S_{\mathrm{eff}}$ in Fig. \ref{lost_rate}. These figures show that the total mass grain rate for $R_{rot}=0$ decreases much more rapidly than those with RATD in the regime where the PR drag dominates over collisions. Therefore, when rotational disruption is absent, and hence the collisional process alone fragments dust less efficiently, the total mass gain rate will become much lower (i.e., the blue curve in Fig. \ref{gain_rate}) due to the predominant loss through the PR drag (i.e., the orange curve in Fig. \ref{lost_rate}).

Comparing Fig. \ref{size_dist} with Fig. \ref{size_distribution_fr}, we find that for $f_{\mathrm{highJ}}<0.8$, the mass in the size range of $\SI{1}{\mu m}$ to $\SI{e3}{\mu m}$ does not change with different values of $S_{\mathrm{eff}}$. This is because, in this size range, the mass is dominated by grains trapped in low-$J$ attractor points. Consequently, the mass in this size range varies with different fractions of grain alignment $f_{\mathrm{highJ}}$.

\subsection{Water Snow Line of the Present Solar System}
Icy grains are particularly interesting because they indicate the location of the snow line. Observations show that C-type asteroids, which are one of the primary source of water-rich chondrites, are located around $\SI{2.7}{au}$ \citep{Morbidelli2000}. However, there is little direct evidence for the existence of surface ice on the asteroids in the main-belt \citep{Dominique2015}, despite the presence of main-belt comets with uncertain compositions \citep{hsieh_population_2006}. Another major source of icy grains in the inner region of the Solar System is comets \cite[e.g.,][]{Mann2017}. \citet{nesvorny2010cometary} found that a significant amount of the zodiacal cloud originates from the disruption of Jupiter family comets.

\citet{min_thermal_2011} shows that the water snow line in the protosolar nebula is sensitive to the model parameters and can not be precisely determined theoretically. Additionally, the effects of planetesimal-formation on the asteroid belt are not well understood \citep{DeMeo2014,Podolak2004}. Therefore, while the exact location of the present water snow line is still debated, it is generally defined as the location where the average temperature falls below $\SI{170}{K}$ and ice grains begin to condense, which is thought to be around $\SI{5}{au}$ \citep{Prockter2005,Henin2018,Dominique2015}. However, for a composite grain with an ice mantle and an effective tensile strength $S_{\mathrm{eff}}$, the present water snow line is also determined by the disruption grain size. In Fig. \ref{disruption_size}, the vertical line indicates the present water snow line, located at $\SI{5}{au}$. The disruption grain size can be calculated by substituting the parameter values at $\SI{5}{au}$ into Equation \ref{a_disr} and by substituting $S_{\mathrm{max}}$ to $S_{\mathrm{eff}}$.
\begin{equation}
    a_{\mathrm{disr,min}}\simeq\num{0.0928}\times S_{\mathrm{eff},9}^{1/5.4}\,\si{.\mu m}.
\end{equation}

Given a specific effective tensile strength, which depends on the structure of dust grains and the thickness of the ice mantle, as shown in Section \ref{Rotational disruption of ice mantles}, we can see from the left panel of Fig. \ref{disruption_size} that large grain sizes would be disrupted by the RAT at the distance beyond the present water snow line $\SI{5}{au}$. Therefore, the number density of icy grains inside the present water snow line defined by RATD is significantly lower than the number density of icy grains that outside the present water snow line as shown in Section \ref{Siz_distribution_result}. Therefore, the snow line for a grain of large size may extend beyond the snow line defined by the sublimation temperature of water ice.

\subsection{Effects of Radiation Pressure on Particle Disruption} \label{Radiation_Pressure}
So far, our results for the RATD effect assumed that grains are on a circular orbit and spun up by radiative torques. Here, we discuss the effect of radiation pressure on the grain motion and the resulting RATD effect.
It is worth mentioning how radiation pressure affects stellar winds and dust grains. For Solar-type stars and late-type stars, \cite{minato_momentum_2006} showed that the stellar wind pressure is ineffective unless $\dot{M}>\num{e3}\dot{M}_\odot$. On the other hand, for early-type stars, the stellar wind is dominated by line-driven winds, which are created by transferring the momentum of radiation to the wind elements via absorption in spectral lines \citep{castor_radiation-driven_1975,lamers_introduction_1999}. The efficiency of the momentum transfer from radiation to wind can be defined as \citep{lamers_introduction_1999}:
\begin{equation}
    \eta_{\mathrm{mom}}=\dot{M}v_{\infty}/(L/c),
\end{equation}
where $\dot{M}$ is the mass loss rate of the star, $v_{\infty}$ is the terminal velocity of the wind, and $L$ is the luminosity of the star. Since most of the momentum of the wind elements is transferred from the momentum of radiation, by momentum conservation, the total pressure after momentum transfer is equal to the radiation pressure from the star, e.g. $P_{\mathrm{tot}}=P_{\mathrm{sw}}+P_{\mathrm{rad}}=\frac{L}{4\pi r^2 c}$. Therefore, the ratio of the ram pressure of the stellar wind $P_{\mathrm{sw}}=\frac{\dot{M}v_{\infty}}{4\pi r^2}$ exerted on a particle to the radiation pressure $P_{\mathrm{rad}}=\frac{L}{4\pi r^2 c}-\frac{\dot{M}v_{\infty}}{4\pi r^2}$ can be estimated as:
\begin{equation}
    \frac{P_{\mathrm{sw}}}{P_{\mathrm{rad}}}=\frac{\eta_{\mathrm{mom}}}{1-\eta_{\mathrm{mom}}}.
\end{equation}
\citet{lamers_introduction_1999} studied the properties of the winds of five typical early-type stars. They found that, except for Wolf–Rayet stars, the ratios $\eta_{\mathrm{mom}}$ for Xi Pup, Epsilon Orionis, P Cygni, and Tau Scorpii are on the order of $0.2$. Therefore, the ram pressure of the stellar wind and the radiation pressure are of the same order for these four stars. The total force exerted on a particle with size $a$ is given by:
\begin{align}
    F_{\mathrm{tot}} &= P_{\mathrm{sw}}\left\langle Q_{\mathrm{sw}}\right\rangle \pi a^2 + P_{\mathrm{rad}}\left\langle Q_{\mathrm{pr}}\right\rangle\pi a^2 \nonumber\\
    &= P_{\mathrm{tot}}\alpha_{\mathrm{rad}}\pi a^2,
\end{align}
where 
\begin{align}
    \alpha_{\mathrm{rad}} &= \frac{P_{\mathrm{sw}}\left\langle Q_{\mathrm{sw}}\right\rangle+P_{\mathrm{rad}}\left\langle Q_{\mathrm{pr}}\right\rangle}{P_{\mathrm{sw}}+P_{\mathrm{rad}}}\nonumber\\
    &= \eta_{\mathrm{mom}}\left\langle Q_{\mathrm{sw}}\right\rangle+(1-\eta_{\mathrm{mom}})\left\langle Q_{\mathrm{pr}}\right\rangle. 
\end{align}
Therefore, the ratios of total force to gravitational force is given by \citep{burns_radiation_1979,hoang_effect_2021}:
\begin{align}
    \beta &= \frac{3L\alpha_{\mathrm{rad}}}{16\pi GMca\rho}\nonumber\\
    &=0.19\frac{L}{L_\odot}\frac{M_\odot}{M}\frac{\alpha_{\mathrm{rad}}}{1}\frac{\SI{3}{g/cm^3}}{\rho}\frac{\SI{1}{\mu m}}{a}, \label{beta_eq}
\end{align} 
where $L/L_\odot$ and $M/M_\odot$ are luminosity and mass of the star in solar units. 

Moreover, the total photogravitational force is given by \citep{krivov2006dust}:
\begin{align} \label{F_pg}
    F_{pg} = -\frac{GM(1-\beta)m}{r^3}\mathbf{r},
\end{align} 
where $G$ is the gravitational constant, $M$ is the mass of the star, and $m$ is the mass of the particle. Therefore, a particle can no longer be held in bound orbits if $\beta>1$, and the released fragments from a parent body on a circular orbit are unbound if $\beta>0.5$ \citep{krivov2006dust}. From Equations \ref{EOM} and \ref{F_pg}, assuming that a particle is released at rest, the derivative of rotational velocity $\omega$ with respect to the distance from the star $r$ is:
\begin{align} \label{dwdr}
    \frac{d\omega}{dr} = \frac{1}{\sqrt{2GM(\beta-1)}}\sqrt{\frac{r_0 r}{(r-r_0)}}\frac {\Gamma_\mathrm{RAT}}{I}\left ( 1- \frac{\omega}{\omega_\mathrm{RAT}}\right ),
\end{align}
where $r_0$ is the initial distance where the particle is released. By mass–luminosity relation, $\beta\propto L \propto M^{\zeta}$, $\Gamma_\mathrm{RAT} \propto L \propto M^{\zeta}$, and $\omega_\mathrm{RAT}\propto L^{1/3} \propto M^{\zeta/3}$. Assuming $\beta \gg 1$ and considering a strong solar radiation field, $\omega_\mathrm{RAT}$ is independent of gas temperature and density to a first-order approximation. Under the same rotational velocity $\omega$ and distance from the star $r$, Equation \ref{dwdr} shows that the change in rotational velocity over the distance from the star is higher for more massive stars if the exponent $\zeta>1$, resulting in faster particle disruption. For stars of approximately solar chemical composition, empirical data show that the exponent $\zeta$ varies between $\num{2.6}$ to $\num{4.5}$ for masses between $\num{\sim 0.2}$ and $\SI{20}{M_\odot}$ \citep{salaris_evolution_2005}. 

\begin{figure}
\includegraphics[width=\columnwidth]{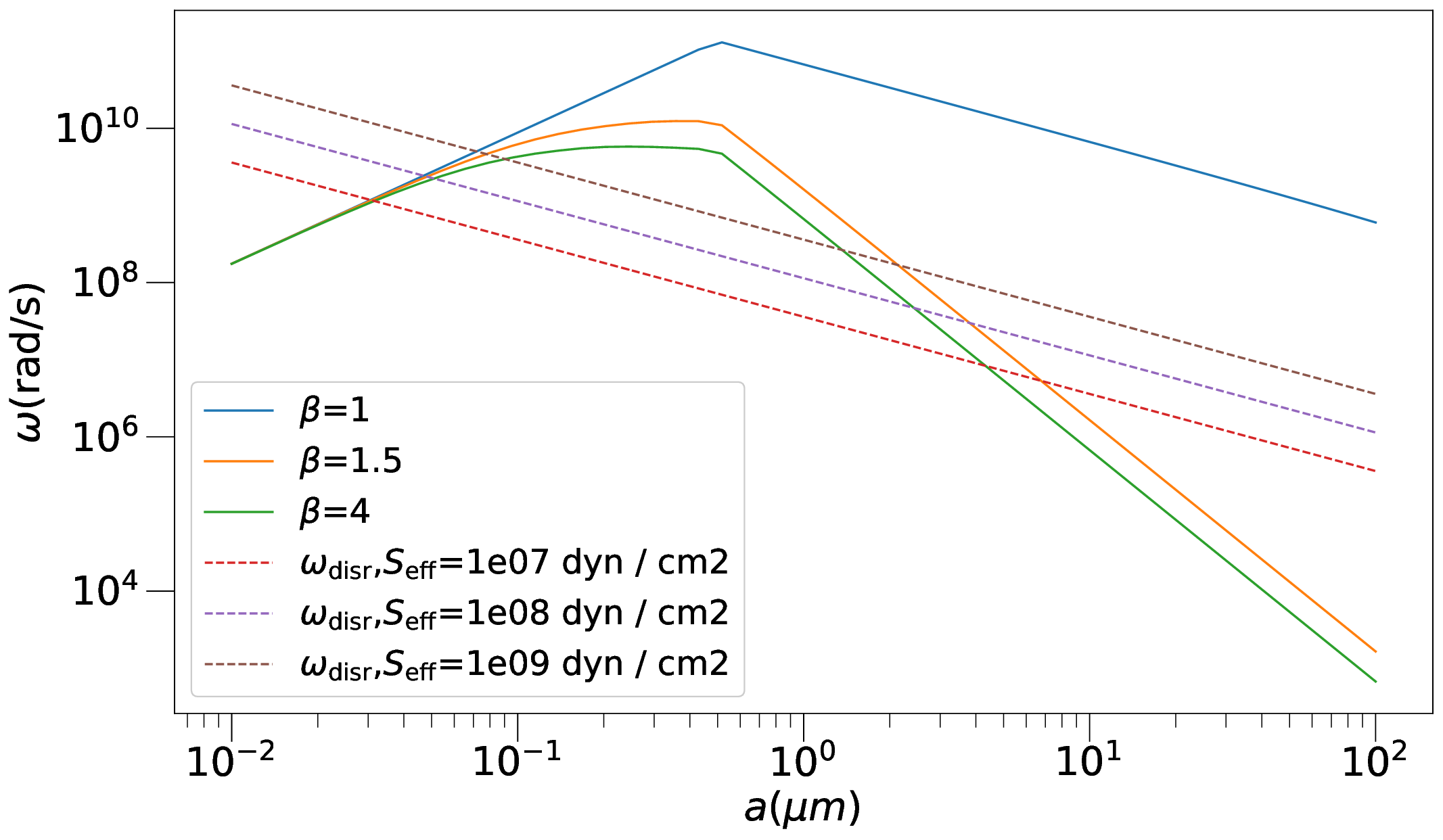}
\caption{Critical rotational velocities ($\omega$, dashed lines) and the maximum rotational velocities ($\omega_\mathrm{disr}$, solid lines) of the particles released at rest at $\SI{1}{au}$ and before reaching $r=\SI{100}{au}$ with different particle sizes ($a$), tensile strengths ($S_{\mathrm{eff}}$) and $\beta$-values. $\beta$ is defined in Equation \ref{beta_eq}.}
\centering \label{Rad_pressure}
\end{figure}

If the $\beta$-value changes due to variations in the parameter $\alpha_{\mathrm{rad}}$, Equation \ref{dwdr} indicates that a larger $\beta$-value results in a smaller change in rotational velocity, making it less likely for the particle to be disrupted at a given distance from the star. Fig. \ref{Rad_pressure} shows the critical rotational velocities $\omega_\mathrm{disr}$ (dashed line) and the maximum rotational velocities $\omega_{\mathrm{max}}$ (solid line) that the particles can achieve as they travel from $\SI{1}{au}$ to $\SI{100}{au}$. These particles are released at rest at $\SI{1}{au}$ with different particle sizes and $\beta$-values. According to Equation \ref{F_pg}, particles with $\beta=1$ will remain at $r=\SI{1}{au}$, where their maximum rotational velocity equals $\omega_\mathrm{RAT}$ at this distance. We observe that particles are expelled by radiation pressure if their $\beta$-value is sufficiently high.

Since radiation pressure depends on the particle properties, \cite{Wilck1996} calculated the $\beta$-value for asteroidal particles, young cometary particles, old cometary particles, and interstellar particles. Their results show that the $\beta$-values for all models are below 1, except for the young cometary particle model. The size of dust grains subject to radiation blowout lies between $\SI{0.02}{\mu m}$ and $\SI{1}{\mu m}$ \citep[see also][]{Editors2009}. This study also implies that some particles in the Solar System can not be blown out by radiation pressure alone, regardless of grain size. This idea is corroborated by \citet{Arnold2019} and \citet{Sterken2019}, who showed the $\beta$-value of different compositions. They found that the $\beta$-value of water ice grains and porous silicate grains in the Solar System is always less than 1, whereas mixtures of water ice, carbon, and silicate can have $\beta$-values greater than 1. However, our results show that RATD effectively destroys particles larger than $\SI{0.1}{\mu m}$ if they are trapped in a high-$J$ attractor point. Thus, dust grains trapped with sizes between $\SI{0.1}{\mu m}$ and $\SI{1}{\mu m}$ are likely subjected to RATD before being blown out by radiation pressure in the Solar System.

\begin{figure}
\includegraphics[width=\columnwidth]{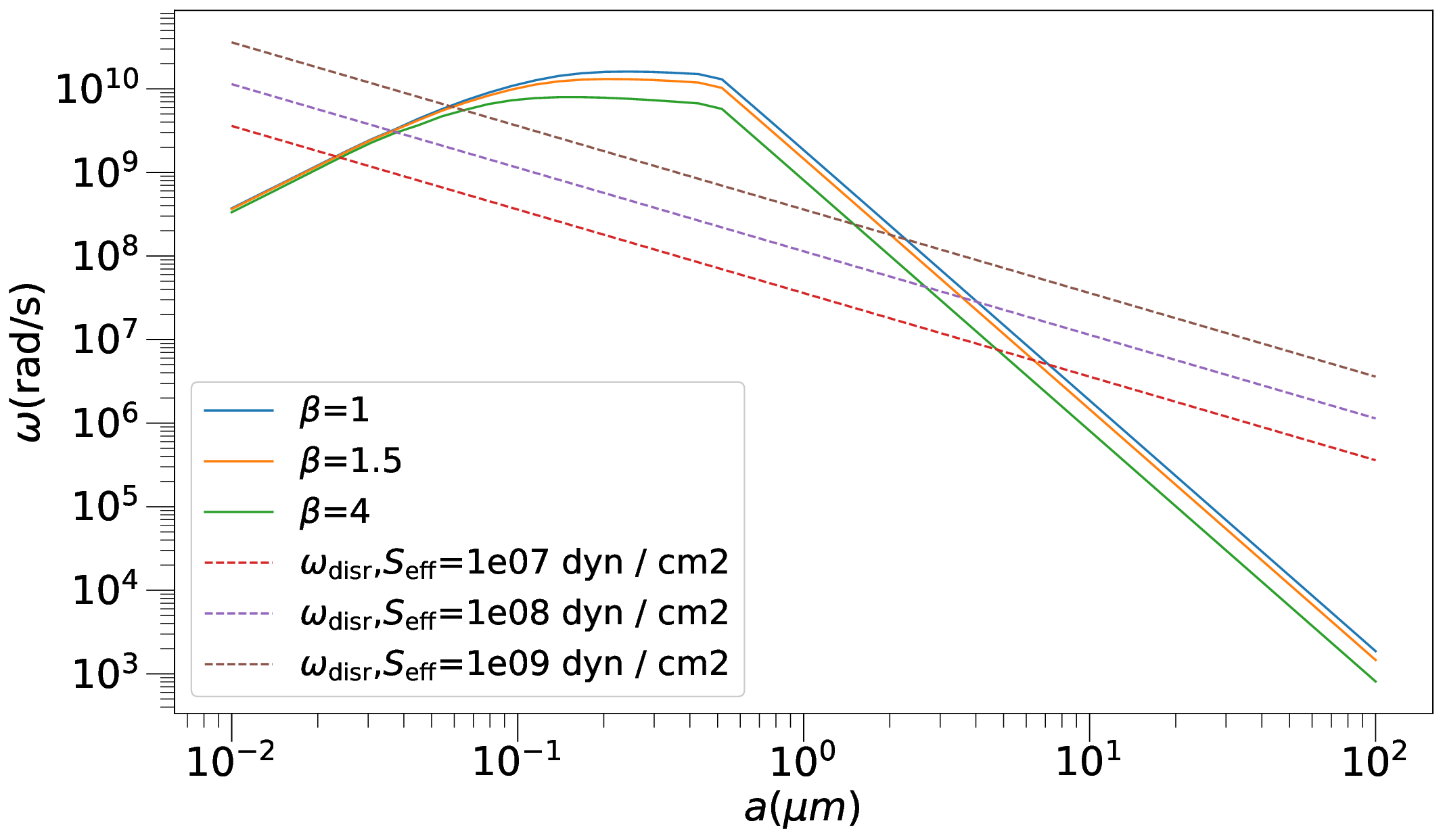}
\caption{Critical rotational velocities ($\omega$, dashed lines) and the maximum rotational velocities ($\omega_\mathrm{disr}$, solid lines) that the $\beta$-meteoroids can reach during the journey between $\SI{1}{au}$ and $\SI{100}{au}$ when released on a circular orbit at $\SI{1}{au}$ with different particle sizes ($a$), tensile strengths ($S_{\mathrm{eff}}$) and $\beta$-values.}
\centering \label{Rad_pressure_beta}
\end{figure}

For $\beta$-meteoroids, here we assume they are released from the parent body at heliocentric distance $r_0$ on circular orbits \citep{Wehry1999}. The radial velocity of the meteoroid is:
\begin{align} \label{vr_cir}
    v_r = \sqrt{-\frac{2GM(\beta-1)}{r}-\frac{l^2}{r^2}+\frac{GM}{r_0}+\frac{2GM(\beta-1)}{r_0}},
\end{align}
where $l=\sqrt{GMr_0}$ is the specific angular momentum of the meteoroid. Therefore, the derivative of rotational velocity $\omega$ with respect to heliocentric distance $r$ is:
\begin{align} \label{dwdr_cir}
    \frac{d\omega}{dr} = \frac{1}{v_r}\frac {\Gamma_\mathrm{RAT}}{I}\left ( 1- \frac{\omega}{\omega_\mathrm{RAT}}\right ).
\end{align}
Fig. \ref{Rad_pressure_beta} shows the critical rotational velocities $\omega_\mathrm{disr}$ (dashed line) and the maximum rotational velocities $\omega_{\mathrm{max}}$ (solid line) that $\beta$-meteoroids can reach during the journey between $\SI{1}{au}$ and $\SI{100}{au}$ when they are released at $\SI{1}{au}$ with different particle sizes and $\beta$-values. The result indicates that $\beta$-meteoroids trapped in a high-$J$ attractor point with sizes between $\SI{0.1}{\mu m}$ to $\SI{1}{\mu m}$ will be disrupted before reaching $\SI{100}{au}$.

\subsection{Location of Disruption in the Presence of Radiation Pressure} \label{Disruption_loc}
\begin{figure}
\includegraphics[width=\columnwidth]{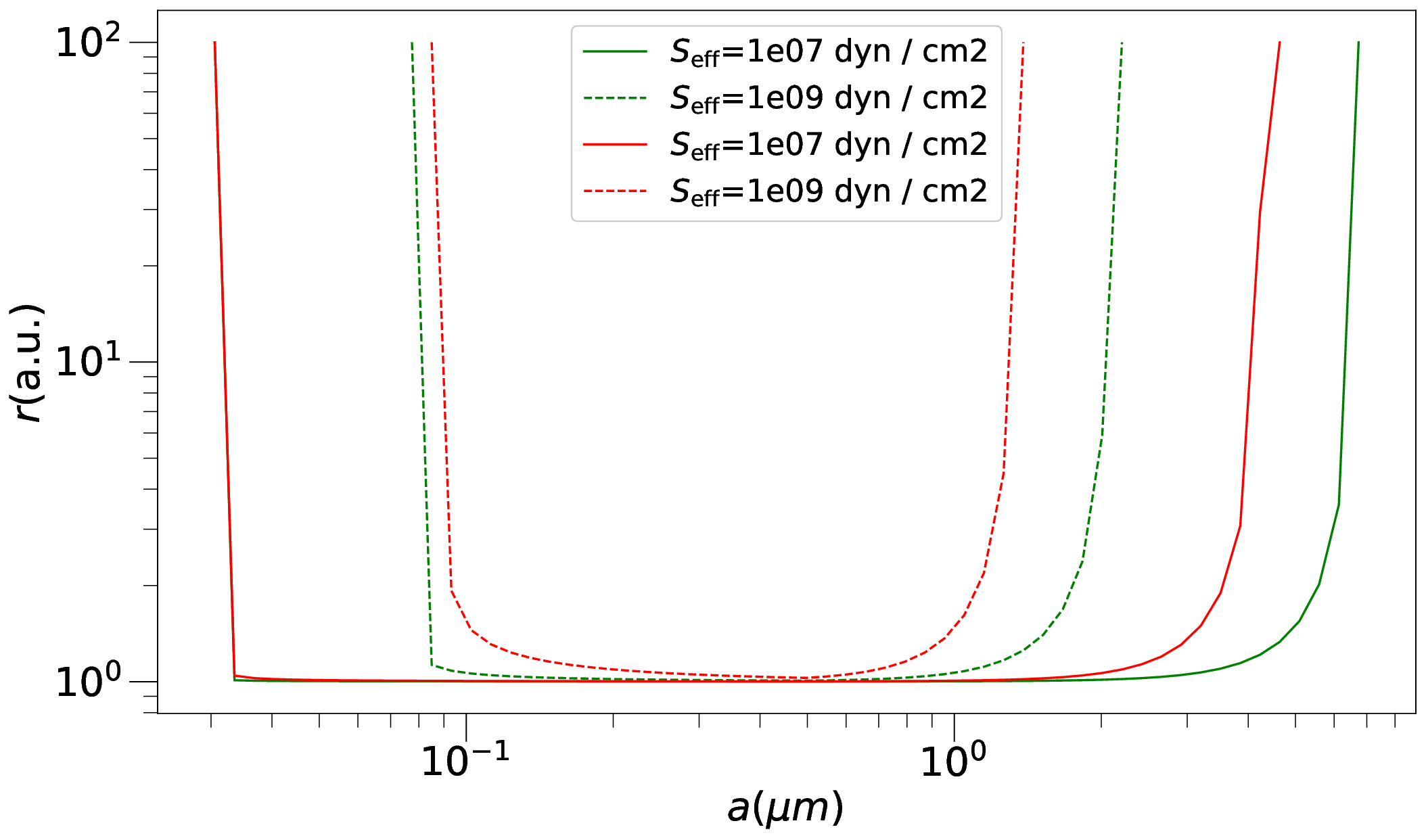}
\caption{The location ($r$) where the particle will be disrupted by RAT when the particle is released at rest at $\SI{1}{au}$ with different particle sizes ($a$), $\beta$-values, and tensile strength ($S_{\mathrm{eff}}$).}
\centering \label{dis_loc_1au}
\end{figure}

\begin{figure}
\includegraphics[width=\columnwidth]{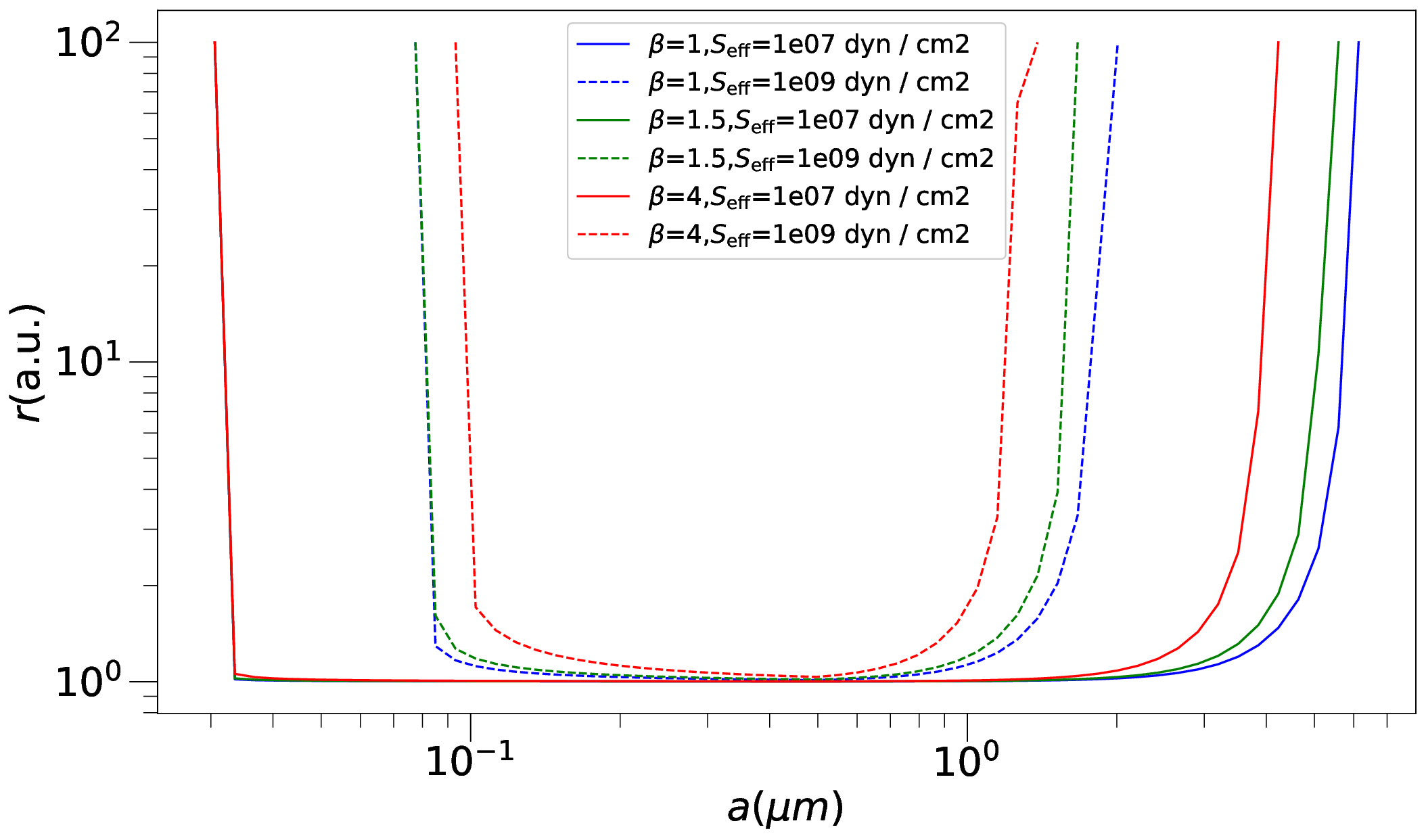}
\caption{The location ($r$) where the particle will be disrupted by RAT when the particle is released with a circular orbit at $\SI{1}{au}$ with different particle sizes ($a$), $\beta$-values, and tensile strength ($S_{\mathrm{eff}}$).}
\centering \label{dis_loc_0.3au}
\end{figure}

Assume the heliocentric distance where the particle is released is disrupted at $r_d$. Then, the displacement interval between the location where the initial distance $r_0$ and the location where the particle is disrupted is $\Delta r=r_d-r_0$. Thus, the location of disruption can be written as:
\begin{align}
    r_d = r_0 + \Delta r.
\end{align}
To estimate the heliocentric distance where the particle is disrupted, we assume that $\Delta r \ll r_0$ so that the acceleration and disruption timescales are approximately constant locally. Therefore, $\Delta r$ can be expressed as:
\begin{align}
    \Delta r \approx v_{r0}t_\mathrm{disr}+\frac{1}{2} \left [\frac{(\beta-1)GM}{r^2}+\frac{l^2}{r^3} \right ]_{r=r_0} t_\mathrm{disr}^2(r_0),
\end{align}
where $v_{r0}$ is the radial velocity of the particle when it is released, and $a_r=\frac{(\beta-1)GM}{r^2}+\frac{l^2}{r^3}$ is the acceleration of the particle in the radial direction. For a particle released at rest, $v_{r0} = 0$ and $l = 0$. By substituting the gravitational constant $G=\frac{4\pi^2}{M_\odot}\frac{\SI{1}{au^3}}{\SI{1}{yr^2}}$, the location of disruption becomes:
\begin{align}
    r_d \approx r_0 + 2\pi^2(\beta-1)\left (\frac{M}{M_\odot}\right )\left (\frac{\SI{1}{au}}{r_0}\right )^2\left ( \frac{t_\mathrm{disr}}{\SI{1}{yr}}\right )^2\,\si{.au}.
\end{align}
Moreover, for a particle released from a circular orbit, $v_{r0} = 0$ and $l=\sqrt{GMr_0}$, the location of disruption is:
\begin{align}
    r_d \approx r_0 + 2\pi^2\beta\left (\frac{M}{M_\odot}\right )\left (\frac{\SI{1}{au}}{r_0}\right )^2\left ( \frac{t_\mathrm{disr}}{\SI{1}{yr}}\right )^2\,\si{.au}.
\end{align}
Therefore, the condition that $\Delta r \ll r_0$ implies that:
\begin{align}
    \left (\frac{M}{M_\odot}\right )\left (\frac{\SI{1}{au}}{r_0}\right )^3\left ( \frac{t_\mathrm{disr}}{\SI{1}{yr}}\right )^2 \ll 1.
\end{align}

The above analysis shows that the displacement interval $\Delta r \propto t_\mathrm{disr}^2/r_0^2$. Furthermore, if the condition $\omega_\mathrm{RAT} \gg \omega_\mathrm{disr}$ is satisfied, by Equation \ref{tdisr01} or Equation \ref{tdisr02} with the same $\beta$-value, the displacement interval is given by:
\begin{equation}
     \Delta r \propto \begin{cases}
     r_0^2 S a^{-1.4}\quad &a\lesssim\bar{\lambdaup}/1.8\\ \label{displacement}
     r_0^2 S a^{4}\quad &a > \bar{\lambdaup}/1.8
     \end{cases}.
\end{equation}
Therefore, the ratio $\Delta r/r_0$ will decrease as $r_0$ decreases, and the condition $\Delta r \ll r_0$ is thus easier to satisfy.

On the other hand, the location of disruption can be calculated numerically by Equation \ref{dwdr_cir} and Equation \ref{omega_disr}. Fig. \ref{dis_loc_1au} and \ref{dis_loc_0.3au} show the heliocentric distance where the particle will be disrupted by RATD, corresponding to the cases in Fig. \ref{Rad_pressure} and \ref{Rad_pressure_beta}, respectively. From those figures, as expected from Equation \ref{displacement}, the displacement interval increases more rapidly for a particle with size $a$ increasing from $\bar{\lambdaup}/1.8$ than decreasing from $\bar{\lambdaup}/1.8$. Additionally, the displacement interval increases as the effective tensile strength $S_{\mathrm{eff}}$ increases. Fig. \ref{dis_loc_1au} and \ref{dis_loc_0.3au} also illustrate that the release location where the particle will be disrupted is very close to the distance it is released for the particle size in the regime $\omega>\omega_\mathrm{disr}$. Furthermore, the slopes of the curves for disruption locations increase steeply at the boundary of the aforementioned regime. Therefore, RATD can effectively destroy the particles before they are blown out by radiation pressure.

\section{Discussion}
\label{discuss}
In this section, we outline the approaches for assessing the RATD theory's impact on dust dynamics within the heliosphere. We also discuss the necessity of in-situ measurements to determine grain size and composition and validate the RATD theory by analyzing dust properties at varying distances from the Sun. Ground-based experiments on samples from missions such as Hayabusa and Stardust are proposed to estimate dust tensile strength and angular velocities, enhancing our understanding of nanoparticle formation. The evolution of debris disks will be discussed, with a focus on how RATD may explain high infrared luminosities observed in systems like HD 113766 by producing nanoparticles at elevated rates. Additionally, we discuss the uncertainties and limitations of our simple models for the investigation of the evolutions of dust size and spatial distributions.

\subsection{Comparison of RAT Efficiency Models and Their Impact on Grain Disruption}

In this paper, we use the parametric function of the RAT efficiency from \citet{Lazarian2007a}, i.e., Equation \ref{RAT_eff}, which was derived by fitting numerical calculations of RATs using DDSCAT \citep{draine_discrete-dipole_2008} for several grain shapes. \cite{herranen_radiative_2019} calculated the RAT efficiency for a large sample of compact grain shapes and found that the mean RAT efficiency is comparable to the model of \cite{Lazarian2007a} (see their Fig. 20). Very recently, \cite{jager_radiative_2024} extended these calculations to include more porous grain structures and found that the RAT efficiency for such shapes is lower than predicted by the \cite{Lazarian2007a} model. However, they also concluded that rotational disruption can still occur when the local radiation field is more than 100 times stronger than the typical interstellar radiation field. Given the much stronger solar radiation field, our results for grain disruption based on the \cite{Lazarian2007a} model remain largely unaffected even if the RAT efficiency for dust aggregates from \cite{jager_radiative_2024} is adopted.

\subsection{In-Situ Dust Measurements and Future Missions}

To verify the location of the present water snow line for different grain sizes, in-situ measurement of the size and composition of the grains is required. Although the distribution of grain size in the Solar System inside $\SI{5}{au}$ has been obtained from multiple measurements (Equation \ref{flux_eq}), how the composition of nanodust changes with radius remains unclear. The highly sensitive dust instrumentation developed in concept missions, such as DuneXpress in the past \citep{grun2009dunexpress}, would allow us to measure the mass and elemental composition simultaneously. Therefore, we can obtain the number density with different tensile strengths and sizes at $\SI{1}{au}$. Moreover, precise size-dependent dust flux measurements provided by DuneXpress would enable us to acquire accurate data to verify the RATD theory at $\SI{1}{au}$ \citep{grun2009dunexpress}. 

Another concept mission, the Interstellar Probe, will also be equipped with dust composition analyzers which are capable of measuring the impactor fluxes for interplanetary and interstellar grains and measuring the composition of IPD and ISD grains \citep{interstellar_probe_study_team_interstellar_2021}. Since the space probe is expected to leave the Solar System, it can measure the mass, speed, and composition of the impacting dust particles at the location of the present water snow line.

\subsection{Ground-Based Measurements}
Laboratory experiments can be conducted on samples collected by in-situ instruments on spacecrafts such as Hayabusa and Stardust spacecrafts. For example, the tensile strength of the ISD can be estimated by measuring the critical angular velocity. \citet{abbas_laboratory_2004} measured the angular velocity of \ch{SiC} particles by analyzing the low-frequency signal of the scattered light. Additionally, a high angular velocity could be measured by Rotational X-ray Tracking (RXT) \citep{Liang2018}. Since RXT has a resolution of tens of microradians, with microsecond time resolution, it can potentially resolve the dust grains with an angular velocity of $\SI{e9}{rad.s^{-1}}$, which is the critical rotational velocity for micrometer-sized dust (Equation \ref{omega_disr}). Confirmation or refutation of this would greatly improve our understanding of nanoparticle formation in the Solar System. 

\subsection{Dust Evolution in Debris Disks}
A debris disk is formed at the end of the protoplanetary disk phase, during which the gas-to-dust mass ratio becomes low. It is usually optically thin, so a strong radiation field is considered. Since the rotational disruption will accelerate the fragmentation process, we would expect that nanoparticles can be produced in young debris disks and in the debris disks around early-type stars at a larger production rate than in the Solar System. Additionally, the number density of dust grains with sizes between $\SI{0.1}{\mu m}$ and $\SI{100}{\mu m}$ would be much lower than the number density produced by grain-grain collisions alone. In contrast, the number density of sub-$\mu m$-sized dust grains will be higher than the number density produced by grain-grain collisions alone.

This might help solve the problem of systems like HD 113766, where the observed infrared fractional luminosity of the inner belt is much higher than the maximum value for a steady state in typical collisional cascades \citep{Su2020}. The outer belt of the system is not affected by RATD because the disk is optically thick to solar wind in the early stage of the planetesimal disk \citep{wetherill_solar_1981}. Consequently, sub-$\mu m$-sized dust grains created in the inner belt are unable to be transported to the outer belt effectively. Additionally, the NIR excess from early-type stars has been modeled as thermal emissions from hot nanoparticles in the vicinity of the stars \citep{Su2013}. 

Moreover, the silicate solid-state features at the wavelength of $\SI{10}{\mu m}$ from some young early-type stars can be attributed to sub-micron particles \citep{Su2013,Su2020}. Later, \cite{rieke_magnetic_2016} showed that nanoparticles can be trapped by a dipole magnetic field. On the other hand, \citet{stamm2019dust} showed that sub-micron-sized particles are expected to be removed in a short time due to the fast radiation blowout of early-type stars with a Parker spiral model. Therefore, they concluded that PR drag can not provide sufficient dust mass to produce the IR excess. Other sources that could contribute nanoparticles, such as comets, were suggested by \citet{Mann2017}. 

Although it might not solve the entire puzzle, our study in Section \ref{Siz_distribution_result} and discussion in \ref{Disruption_loc} provide a mechanism where the even faster RATD process might be able to avoid the removal barriers of micron-sized particles by swiftly breaking them down to nanoparticles, thus increasing the number of nanoparticles. In reality, it is highly intriguing to consider the interplay between collisional cascade, PR drag, stellar winds, RATD, the presence of a gas component \citep[e.g.,][]{Lebreton2013}, and impact events \citep[e.g.,][]{Su2020} for nanoparticles production and loss. Modeling these complex processes is beyond the scope of this paper.  

\subsection{Limitations of Our Theoretical Investigation}
For larger particles ($>\SI{e2}{\mu m}$), the numerical results in Fig. \ref{Number} indicate that the slope of the size distribution at $\fidloc$ differs from the in-situ measurement from \citet{Grun1985}. The discrepancy could be attributed to the limited sources of larger particles (i.e., which is modelled using one free parameter $\dot{m}^{-}_\mathrm{b}$ in our simple analysis) and the influence of the Earth's gravitational field that sweeps IDPs \citep[e.g.,][]{jorgensen_distribution_2021}.

Conversely, while models based on infrared emission from the zodiacal dust cloud provide the spatial distribution of micrometer-sized dust \citep{rowan2013improved}, sub-$\mu m$-sized dust grains are unlikely to follow the same distribution. These smaller grains, with their high charge-to-mass ratios, are significantly affected by magnetic forces. Therefore, they can be captured and carried away by solar winds \citep[e.g.,][]{Czechowski2010,juhasz2013dynamics,obrien_development_2014,obrien_effects_2018}. Consequently, as noted in Section \ref{RATD}, for these charged nanodust, their high-speed trajectories are preferentially detected by plasma wave instruments and are not modeled in the present study. Moreover, because the Sun's magnetic field fluctuates over time, these particles are expected to be displaced from the zodiacal plane \citep[e.g.,][]{chiang_stellar_2017,mukai_modification_1984}. As a result, the number density of dust grains decreases, leading to a lower mass loss rate from grain-grain collisions. Besides magnetic effects, solar winds may destruct small grains through nonthermal sputtering \citep{hoang_effect_2021}.

For $\beta$-meteoroids that trapped in a low-$J$ attractor, they will not be disrupted by RATD and PR drag, and will only be disrupted by collisions with other grains. However, given the large timescale for collision disruption shown in Fig. \ref{characteristic_time}, these $\beta$-meteoroids will likely be pushed outward from the heliosphere into interstellar space before they experience collisions. 

While more sophisticated modeling is reserved for future studies, it is important to highlight that, although the loss rate of sub-$\mu m$-sized dust grains remains uncertain, their production rate can be calculated as detailed in Section \ref{sd_intro}, where RATD is considered in a dust collisonal cascade model for the first time. This information is essential for guiding future research.

\section{Summary}
\label{summary}

We have investigated the lifetime of dust grains in the heliosphere by applying the RATD mechanism and have incorporated this mechanism into the dust size distribution model proposed by \citet{wyatt_debris_2011}. We have also studied the effect of RATD on the dust strongly subject to radiation pressure. Our main results are summarized as follows:

\begin{enumerate}
\item We have calculated the characteristic disruption timescales due to radiative torque for various tensile strengths of dust. Our results show that large particles (ranging from >$\SI{0.1}{\mu m}$ to <$\SI{e3}{\mu m}$)) can indeed be fragmented into nanoparticles (less than $\SI{50}{nm}$) within the heliosphere (less than $\SI{100}{au}$), regardless of their distance from the Sun.

\item The LISM affects the rotational disruption size and time in the outer heliosphere (i.e., $>\SI{10}{au}$). Specifically, it significantly impacts the maximum disruption size while slightly influencing the minimum disruption size. Therefore, we conjecture that micrometer-sized dust can be fractured into nanoparticles by RATD within the entire heliosphere. RATD plays an important role in shaping the size distribution of $\mu m$-sized dust.

\item We have calculated the steady-state dust size distribution using a model based on the work of \cite{wyatt_debris_2011}. Our results indicate that rotational disruption is the most effective mechanism for breaking apart dust particles with sizes ranging from $\SI{0.1}{\mu m}$ to $\SI{e3}{\mu m}$. Additionally, we compared the size distributions for different fractions of grain alignment. The comparison reveals that size distributions with $f_{\mathrm{highJ}}<0.8$ closely resemble those where rotational disruption is absent, in contrast to the distributions observed when $f_{\mathrm{highJ}}=1$.

\item We have calculated the present locations of the water snow line for different grain sizes and found that the snow line of large dust grains can exceed the limit imposed by the sublimation temperature of water ice.

\item We have shown that for a dust grain with a constant $\beta$-value, increasing the star's luminosity will not cause the particle to be expelled if the dust remains trapped in a high-$J$ attractor. In this case, the particle is disrupted more rapidly by radiative torque compared to being blown out by radiation pressure. Therefore, detected $\beta$-meteoroids should be trapped in a low-$J$ attractor in our model.
\end{enumerate}

\section*{Acknowledgement}
We sincerely thank the reviewers for their helpful comments and suggestions, which significantly improved the presentation of this paper. This work is supported by the NSTC grant in Taiwan through NSTC 113-2112-M-001-012. T.H. acknowledge the support from the main research project (No. 2025186902) from Korea Astronomy and Space Science Institute (KASI).  

\section*{Data availability}
The data underlying this article will be shared on reasonable request to the corresponding author.



\bibliographystyle{mnras}
\bibliography{main} 




\appendix

\section{Characteristic Damping Time} \label{Appendix A}
Assume that the gas reaches thermal equilibrium, so the gas thermal velocity follows the Boltzmann distribution with a constant drift velocity $v_\mathrm{d}$, given by:
\begin{equation}
    f(\pmb{v})=\left(\frac{m}{2\pi kT}\right )^{3/2} e^{-m(\pmb{v}-\pmb{v}_d)^2/2kT}.
\end{equation}
We also assume the dust particles are spherical, with the surface defined by spherical coordinates $(a,\theta',\phi')$, where $a$ is the radius of the sphere. Without loss of generality, assume the direction of the drift velocity along $\theta'=0$.

\begin{figure}
\includegraphics[width=\columnwidth]{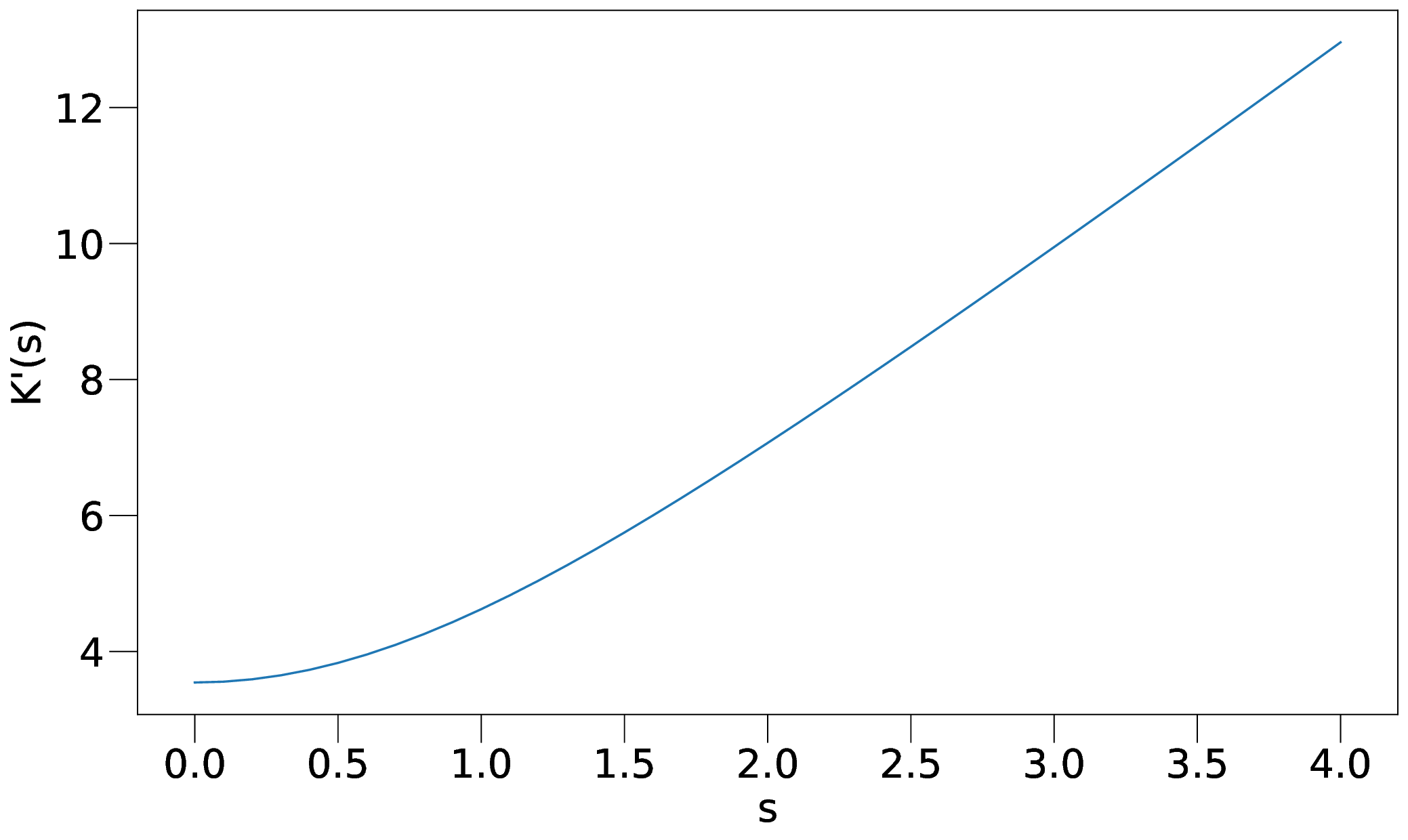}
\caption{Numerical result of the integral function $K'(s)$ as a function of the velocity ratio $s=v_\mathrm{d}/v_\mathrm{th}$, where $v_\mathrm{d}$ is the drift velocity and $v_\mathrm{th}$ is the thermal gas velocity.}
\centering \label{A1}
\end{figure}

\begin{figure}
\includegraphics[width=\columnwidth]{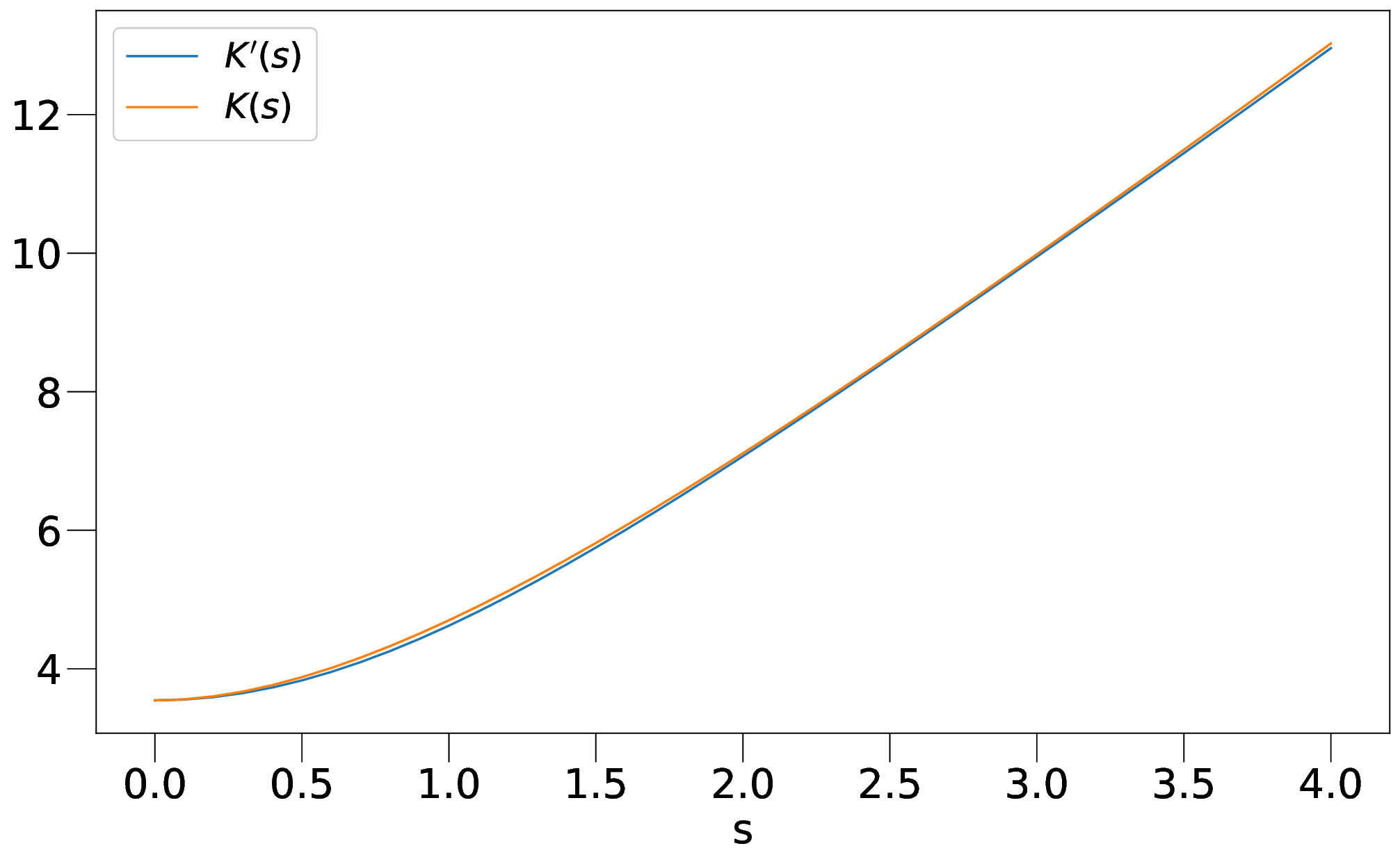}
\caption{Comparison of the numerical result $K'(s)$ and the approximated result $K(s)$ of the integral function as a function of the velocity ratio $s=v_\mathrm{d}/v_\mathrm{th}$, where $v_\mathrm{d}$ is the drift velocity and $v_\mathrm{th}$ is the thermal gas velocity.}
\centering \label{A2}
\end{figure}

The number of particles passing through the area $dA$ with the speeds in the interval $[v,v+dv]$ over the time $dt$ is given by:
\begin{equation}
\begin{aligned}
    \frac{dN}{dt}&=n\int_0^{\pi}\int_0^{\infty}\int_0^{2\pi}\int_0^{\frac{\pi}{2}}v\cos{\theta}f(\pmb{v})d\theta d\phi v^2\sin{\theta}dv\times \\
    &2\pi a^2\sin{\theta'}d\theta',
\end{aligned}
\end{equation}
\begin{equation}
\begin{aligned}
    \frac{dN}{dt}&=n\int_0^{\pi}\int_0^{\infty}\int_0^{2\pi}\int_0^{\frac{\pi}{2}} v^3\sin{\theta}\cos{\theta}\left(\frac{1}{v_\mathrm{th}^2\pi}\right)^{3/2}\times \\
    &e^{-(v^2-2vv_\mathrm{d}\sin{\theta}\sin{\theta'}\cos{\phi}-2vv_\mathrm{d}\cos{\theta}\cos{\theta'}+v_\mathrm{d}^2)/v_\mathrm{th}^2}\times \\
    &d\theta d\phi dv 2\pi a^2\sin{\theta'}d\theta'.
\end{aligned}
\end{equation}
Let $x=\frac{v}{v_\mathrm{th}}$,$s=\frac{v_\mathrm{d}}{v_\mathrm{th}}$
\begin{equation}
\begin{aligned}
    \frac{dN}{dt}&=na^2v_\mathrm{th}\int_0^{\pi}\int_0^{\infty}\int_0^{2\pi}\int_0^{\frac{\pi}{2}} x^3\sin{\theta}\cos{\theta}\left(\frac{1}{\pi}\right)^{3/2}\times \\
    &e^{-(x^2-2xs\sin{\theta}\sin{\theta'}\cos{\phi}-2xs\cos{\theta}\cos{\theta'}+s^2)}\times \\
    &d\theta d\phi dx 2\pi \sin{\theta'}d\theta'.
\end{aligned}
\end{equation}
We can define the integral function $K(s)$ as:
\begin{equation}
\begin{aligned}
    K'(s)&=\int_0^{\pi}\int_0^{\infty}\int_0^{2\pi}\int_0^{\frac{\pi}{2}} x^3\sin{\theta}\cos{\theta}\left(\frac{1}{\pi}\right)^{3/2}\times \\
    &e^{-(x^2-2xs\sin{\theta}\sin{\theta'}\cos{\phi}-2xs\cos{\theta}\cos{\theta'}+s^2)}\times \\
    &d\theta d\phi dx 2\pi \sin{\theta'}d\theta',
\end{aligned}
\end{equation}
whose numerical result is shown in Fig.\ref{A1}
The result can be approximated as:
\begin{equation}
    K(s)=\pi(s+e^{-s}-1)+2\sqrt{\pi}.
\end{equation}
Fig. \ref{A2} illustrates the good agreement between the numerical and the approximated result.

Therefore, the damping timescale of a dust grain damped by hydrogen gas can be approximated as:
\begin{equation}
    \tau_\mathrm{H} =\frac{I}{\frac{2}{3}m_\mathrm{H}a^2\frac{dN}{dt}}= \frac{3I}{2 n_\mathrm{H} m_\mathrm{H} a^4 v_\mathrm{th}K(s)}.
\end{equation}
In the case where $v_\mathrm{d}\gg v_\mathrm{th}$, $K(s)\approx \pi s$, we recover the result in \citet{Hoang2019_2} as:
\begin{equation}
    \tau_\mathrm{H}(v_\mathrm{d}) = \frac{3I}{2n_\mathrm{H} m_\mathrm{H} \pi a^4v_\mathrm{d}}.
\end{equation}
For the case where $v_\mathrm{d}\ll v_\mathrm{th}$, $K(s)\approx 2\sqrt{\pi}$, we recover the result in \citet{Hoang2019} as:
\begin{equation}
    \tau_\mathrm{H}(v_\mathrm{d}) = \frac{3I}{4\sqrt{\pi}n_\mathrm{H} m_\mathrm{H} a^4v_\mathrm{th}}.
\end{equation}

\section{Radiation Pressure Cross-Section Efficiency in the Rayleigh Regime} \label{Appendix B}
The radiation pressure efficiency, as described by \citep{voshchinnikov1990radiation}, is:
\begin{align}
   Q_{\mathrm{pr}} = Q_{abs} + (1-\left\langle \cos{\Theta}\right\rangle)Q_{sca},
\end{align}
where $\Theta$ is the angle between the incident and scattered radiation, and $\left\langle \cos{\Theta}\right\rangle$ is the mean value of $\cos{\Theta}$.
For particles in the Rayleigh regime, with size parameter $x=2\pi a/\lambdaup$, the absorption efficiency is given by \citep{friedlander_smoke_2000} as:
\begin{align}
   Q_{abs} = -4x\Im{\frac{m^2-1}{m^2+2}},
\end{align}
and the scattering efficiency is:
\begin{align}
   Q_{sca} = \frac{8}{3}x^4 \Re{\frac{m^2-1}{m^2+2}}^2,
\end{align}
where $m$ is the refractive index of the particle, which depends on the material of the particle and the wavelength of the incident radiation. Since $Q_{sca}$ is much smaller compared to $Q_{abs}$, the radiation pressure efficiency can be approximated as:
\begin{align}
   Q_{\mathrm{pr}} =& -4x\Im{\frac{m^2-1}{m^2+2}}  + \frac{8}{3}x^4 \Re{\frac{m^2-1}{m^2+1}}^2(1-\left\langle \cos{\Theta}\right\rangle) \nonumber\\
   \simeq& -4x\Im{\frac{m^2-1}{m^2+2}}. \label{Qpr_R}
\end{align}
Equation \ref{Qpr_R} shows that the radiation pressure efficiency in the Rayleigh regime is $Q_{\mathrm{pr}} \propto a$. The average radiation pressure efficiency is:
\begin{align}
    \left\langle Q_{\mathrm{pr}} \right\rangle =& \frac{\int{Q_{\mathrm{pr}} u_\lambdaup d\lambdaup}}{\int{u_\lambdaup d\lambdaup}}.
\end{align}

\section{Effects of Interplanetary Magnetic Field on Grain Alignment and Disruption} \label{Appendix C}
In general, in the absence of magnetic fields, dust grains are aligned solely by radiative torques, where the angular momentum vector $\pmb{J}$ aligns with the radiation direction ($\pmb{k}$). This alignment mechanism is referred to as $k$-RAT in the context of grain alignment physics \citep{hoang_internal_2023}. A fraction of grains, denoted by $f_\mathrm{highJ}$, achieve alignment at high-$J$ attractor points under this mechanism. In the presence of magnetic torques, the grain undergoes Larmor precession, causing $\pmb{J}$ to align with the external magnetic field. This alignment process is termed $B$-RAT \citep{hoang_internal_2023}. Magnetic torques play a crucial role in aligning grains with the magnetic field, and the fraction $f_\mathrm{highJ}$ differs from that observed in $k$-RAT. In the presence of an electric torque, the grain will precess around the electric field $E$, which is referred to as $E$-RAT. The detailed discussion of these processes can be found in \citep{hoang_internal_2023}. Below, we briefly discuss the effects of interplanetary magnetic fields on the grain rotational dynamics.

\subsubsection{Interplanetary magnetic fields}
Inside the heliosphere, the large-scale structure and dynamics of the heliospheric magnetic field are mainly shaped by the solar wind flow \citep{owens_heliospheric_2013}. \citet{Lazarian2007a} demonstrated that dust grains can be trapped in high-$J$ attractor points in the presence of a magnetic field due to the fast Larmor precession. However, if the dust grain's center of charge is offset from its center of mass, the Lorentz force exerts a torque on it. To assess the magnitude of this torque, it is essential to evaluate the strengths of both the magnetic field and the electric dipole moment.

The radial component of the magnetic field in the heliosphere can be approximated using an empirical formula derived from in situ measurements \citep{mann_heliospheric_2023}:
\begin{equation}
    B_r(r) = \frac{6}{r^3}R^3_\odot + \frac{1.18}{r^2}R^2_\odot,
\end{equation}
where $r$ is the heliocentric distance, $R_\odot$ is the radius of the Sun, and $B_\mathrm{r}$ is given in Gauss (G). The azimuthal component of the magnetic field in the ecliptic plane can then be determined using the Parker model \citep{owens_heliospheric_2013}:
\begin{equation}
    B_\phi(r) = -B_r(r) \cdot \frac{\Omega_\odot r}{v_\mathrm{p}},
\end{equation}
where $\Omega_\odot = \SI{2.968e-6}{s^{-1}}$ is the angular velocity of the Sun, and the speed of the solar wind $v_\mathrm{p}$ can be obtained from the following equation \citep{mann_heliospheric_2023}:
\begin{equation}
    v'(r)^2 - \ln{v'(r)^2} = 4 \cdot\ln{r'} + \frac{4}{r'} - 3,
\end{equation}
with $v'(r) = v_\mathrm{p}(r)/v_c$ (where $v_c = (kT/\Tilde{\mu} m_\mathrm{p})^{1/2}$ is the critical velocity) and $r' = r / r_c$ (where $r_c = GM_{\odot}/2v_c^2$ is the critical radius). Here, $k$ is the Boltzmann constant, $T = \SI{8.3e4}{K}$ is the temperature of the solar wind, $m_\mathrm{p}$ is the proton mass, $G$ is the gravitational constant, $M_{\odot}$ is the mass of the Sun, and $\Tilde{\mu} = 0.57$ is the mean molecular weight of the heliospheric plasma.

Therefore, the magnitude of the magnetic field $B_\mathrm{tot}$ can be expressed as:
\begin{align}
    \abs{\pmb{B}_\mathrm{tot}(r)} &= \abs{B_r(r)\pmb{\hat{r}} + B_\phi(r)\pmb{\hat{\phi}}} \nonumber \\
    &= B_r(r)\sqrt{1+\left (\frac{\Omega_\odot r}{v_\mathrm{p}}\right )^2}.
\end{align}

\subsubsection{Interaction of magnetic fields with grain magnetic dipole moment}
Now, we examine the effects of interplanetary magnetic field on grain rotational dynamics, which are governed by the interaction between the magnetic field and grain electric or magnetic dipole moments. 

We first discuss the interaction of magnetic fields with the grain magnetic dipole moments. When a charged dust grain rotates, a magnetic dipole moment is generated due to the motion of charges and the Barnett effect. The torque exerted by the magnetic field on the magnetic dipole causes precession around the magnetic field direction. Let us assume that a dust grain of size $a$ rotates with an angular velocity $\omega$. The magnitudes of the magnetic dipole moment $\abs{\pmb{\mu}_b}$ and the corresponding torque $\abs{\Gamma_b}$ can be estimated by modeling the grain as a uniformly charged sphere with charge $Q = a V$ \citep{weingartner_disalignment_2006}:
\begin{align}
    \abs{\pmb{\mu}_b} &= \frac{1}{3}a^2 Q \omega \nonumber \\
    &\approx \num{1.1e-19}\left (\frac{V}{\SI{3}{V}}\right )\left (\frac{a}{\SI{0.1}{\mu m}}\right )^3\left (\frac{\omega}{\SI{e9}{s^{-1}}}\right )\,\si{.dyne.cm.G^{-1}},
\end{align}
\begin{align} \label{Tau_b}
    \abs{\pmb{\Gamma}_b} &= \abs{\pmb{\mu}_b \pmb{\times} \pmb{B}_\mathrm{tot}} \nonumber \\
    &= \abs{\pmb{\mu}_b} B_r(r)\sqrt{1+\left (\frac{\Omega_\odot r}{v_\mathrm{p}}\right )^2}\abs{\sin{\phi_b}},
\end{align}
where $\phi_b$ is the angle between the magnetic dipole vector $\pmb{\mu}_b$ and the magnetic field $\pmb{B}_\mathrm{tot}$, $V$ is the surface potential of heliospheric dust \citep{horanyi_charged_1996,kimura_electric_1998}, and $\SI{e9}{s^{-1}}$ is the fastest spin rate of a micron-sized dust before rotational disruption (see Equation \ref{omega_disr}).

The magnetic dipole moment can also be generated by Barnett effect. To evaluate its magnitude, consider silicate grains with the structural unit \ch{MgFeSiO4} for paramagnetic dust grains, and carbonaceous grains with $\approx$10\% H incorporation for weakly paramagnetic dust grains. The magnitude of the magnetic dipole moments $\abs{\mu_\mathrm{Bar}}$ are given by \citep{weingartner_disalignment_2006}:
\begin{align}
    \abs{\pmb{\mu}_\mathrm{Bar}}\text{(sil)} &\approx \num{1.2e-15}\left (\frac{T_\mathrm{d}}{\SI{15}{K}}\right )^{-1}\left (\frac{a}{\SI{0.1}{\mu m}}\right )^3 \nonumber \\
    &\times \left (\frac{\omega}{\SI{e9}{s^{-1}}}\right )\,\si{.dyne.cm.G^{-1}},
\end{align}
for silicate grains, and
\begin{align}
    \abs{\pmb{\mu}_\mathrm{Bar}}\text{(carb)} &\approx \num{1.5e-19}\left (\frac{T_\mathrm{d}}{\SI{15}{K}}\right )^{-1}\left (\frac{a}{\SI{0.1}{\mu m}}\right )^3 \nonumber \\
    &\times \left (\frac{\omega}{\SI{e9}{s^{-1}}}\right )\,\si{.dyne.cm.G^{-1}},
\end{align}
for carbonaceous grains. The magnitude of the corresponding torque $\abs{\Gamma_\mathrm{Bar}}$ is given by:
\begin{align} \label{Tau_bar}
    \abs{\pmb{\Gamma}_\mathrm{Bar}} &= \abs{\pmb{\mu}_\mathrm{Bar} \pmb{\times} \pmb{B}_\mathrm{tot}} \nonumber \\
    &= \abs{\pmb{\mu}_\mathrm{Bar}} B_r(r)\sqrt{1+\left (\frac{\Omega_\odot r}{v_\mathrm{p}}\right )^2}\abs{\sin{\phi_\mathrm{Bar}}},
\end{align}
where $\phi_\mathrm{Bar}$ is the angle between the magnetic dipole vector $\pmb{\mu}_\mathrm{Bar}$ and the magnetic field $\pmb{B}_\mathrm{tot}$.

Due to these magnetic torques, the grain undergoes Larmor precession, causing $\pmb{J}$ to align with the external magnetic field via B-RAT (see e.g., \citealt{hoang_internal_2023}).
 
\subsubsection{Interaction with grain electric dipole moment}
A dust grain of asymmetric charge distribution has an electric dipole moment, which can be described by $\pmb{\mu}_{\mathrm{e}}= Q\cdot\xi a \pmb{\hat{x}}$, where $\xi a$ represents the displacement between the center of mass of the grain and the centroid of the charge \citep{DraineLazarian1998}, $\hat{x}$ is the unit vector pointing from the center of mass to the centroid of the charge, and $Q = a V$ represents the total charge on the grain. Here, $V = \SI{3}{V}$ is the surface potential \citep{kimura_electric_1998}.

When a dust grain moves across the magnetic field, it experiences an induced electric field, $\pmb{E}_\mathrm{ind} = \left (\frac{\pmb{v}}{c}\right ) \pmb{\times} \pmb{B}$, due to the interplanetary magnetic field \citep{hoang_electromagnetic_2017}. The torque exerted by this induced electric field on the grain's electric dipole moment ($\pmb{\Gamma}_\mathrm{ed}$) can induce the precession of the grain angular momentum around $\pmb{E}$, which can lead to the alignment of $J$ with $\pmb{E}$ by RATs, which is called $E$-RAT (e.g., \citealt{lazarian_two_2020}).

The magnitude of this electric torque can be calculated as \citep{hoang_electromagnetic_2017,lhotka_kinematic_2019}:
\begin{align} \label{Tau_ed}
    \abs{\pmb{\Gamma}_\mathrm{ed}} &= \abs{\pmb{\mu}_e \pmb{\times} \pmb{E}_\mathrm{ind}} \nonumber \\
    &\simeq \xi a^2 V\left (\sqrt{\frac{GM_\odot}{r}}-\Omega_\odot r \right )\cdot B_r(r) \abs{\sin{\phi_\mathrm{ed}}},
\end{align}
where $\Omega_\odot = \SI{2.968e-6}{s^{-1}}$ is the angular velocity of the Sun, and $\phi_\mathrm{ed}$ is the angle between the electric dipole vector $\pmb{\mu}_e$ and the cross-product vector $\pmb{v} \pmb{\times} \pmb{B}$. 
 
For dust grains with high angular momentum $J$, the net electric and magnetic dipole moments, averaged over rapid spinning, are aligned with $J$. As a result, the net torque acts perpendicular to $J$, leaving the magnitude of the angular velocity unchanged \citep{hoang_internal_2023}. Conversely, for dust grains with low angular momentum $J$, the electric dipole torque can align with $J$ and slow the grain’s rotation. Assuming a dust grain with $a > \bar{\lambdaup}/1.8$ and $r \gg \sqrt[3]{\frac{GM_\odot}{\Omega_\odot^2}} \approx \SI{0.17}{au}$, the first term in Equation \ref{Tau_ed} becomes negligible. The heliocentric distance $r_a$, at which $\Gamma_\mathrm{RAT} = \abs{\pmb{\Gamma}_\mathrm{ed} (\phi_\mathrm{ed} = \pi/2)}$, can be derived by combining Equations \ref{tauR} and \ref{Tau_ed}:
\begin{equation}
    r_a \approx \frac{2.22}{\xi}\,\si{.au}.
\end{equation}
For $r < r_a$, $\Gamma_\mathrm{RAT} > \abs{\pmb{\Gamma}_\mathrm{ed} (\phi_\mathrm{ed} = \pi/2)}$, resulting in continuous spin-up of the dust grain. Using $\xi = 0.01$ as suggested by \citet{DraineLazarian1998}, we find $r_a \approx \SI{222}{au}$. This calculation indicates that within the heliosphere, radiative torques consistently dominate over electric dipole torques. Consequently, for simplicity, we neglect the effects of electric torques in the discussion of RATD for the remainder of this paper.

\subsubsection{Relative effects of magnetic to electric torques} \label{mtoe}
Due to the electric torques, the grain’s electric dipole moment, and consequently $\pmb{J}$, can precess around the induced electric field $\pmb{E}_\mathrm{ind}$. This results in $\pmb{J}$ aligning with $\pmb{E}_\mathrm{ind}$, a process known as $E$-RAT \citep{hoang_internal_2023}. The fraction $f_\mathrm{highJ}$ in this scenario also differs from the fractions observed in $k$-RAT and $B$-RAT.

When both electric and magnetic torques act on a dust grain, its precession behavior depends on the relative magnitudes of the torques. If the magnetic dipole torque $\abs{\pmb{\Gamma}_b}$ or the torque induced by the Barnett effect $\abs{\pmb{\Gamma}_\mathrm{Bar}}$ dominates over the electric dipole torque $\abs{\pmb{\Gamma}_\mathrm{ed}}$, the grain precesses around the magnetic field $\pmb{B}$. Conversely, if the electric dipole torque is dominant, precession occurs around the induced electric field $\pmb{E}_\mathrm{ind}$. For clarity, let $\abs{\pmb{\Gamma}_\mathrm{ed,max}}$, $\abs{\pmb{\Gamma}_{b,\mathrm{max}}}$, and $\abs{\pmb{\Gamma}_\mathrm{Bar,max}}$ represent the maximum torques corresponding to $\phi_\mathrm{ed} = \pi/2$, $\phi_b = \pi/2$, and $\phi_\mathrm{Bar} = \pi/2$, respectively. 

To better understand the relative importance of magnetic, electric, and radiative torques on grain alignment, we need to compare the characteristic timescales of precession induced by these torques, as in e.g., \cite{hoang_internal_2023}. Here we compare their relative magnitudes for convenience. The ratio of magnitude between the electric dipole torque, $\abs{\pmb{\Gamma}_\mathrm{ed,max}}$, and the magnetic dipole torque, $\abs{\pmb{\Gamma}_{b,\mathrm{max}}}$. This ratio can be derived by combining Equations \ref{Tau_ed} and \ref{Tau_b}:
\begin{align} \label{r_ed_b}
    \frac{\abs{\pmb{\Gamma}_\mathrm{ed,max}}}{\abs{\pmb{\Gamma}_{b,\mathrm{max}}}} \approx& \frac{3 \xi \Omega_\odot r}{\omega a} \left(1+\left (\frac{\Omega_\odot r}{v_\mathrm{p}}\right)^2\right)^{-\frac{1}{2}} \nonumber \\
    \approx& \num{133}\left (\frac{a}{\SI{0.1}{\mu m}}\right )^{-1}\left (\frac{\omega}{\SI{e9}{s^{-1}}}\right )^{-1}\left (\frac{r}{\SI{1}{au}}\right ) \nonumber \\
    &\times \left(1+\left (\frac{\Omega_\odot r}{v_\mathrm{p}}\right)^2\right )^{-\frac{1}{2}}.
\end{align}
For a dust grain trapped at high-$J$ attractor points, the fastest spin rate is determined by the critical rotational velocity $\omega = \omega_\mathrm{disr}$ (see Equation \ref{omega_disr}). By combining Equations \ref{omega_disr} and \ref{Tau_b}, we derive:
\begin{align} \label{r_eb_hJ}
    \frac{\abs{\pmb{\Gamma}_\mathrm{ed,max}}}{\abs{\pmb{\Gamma}_{b,\mathrm{max}}}} \approx \num{37}\hat{\rho}^{1/2}S_{\mathrm{max},9}^{-1/2} \left (\frac{r}{\SI{1}{au}}\right ) \left(1+\left (\frac{\Omega_\odot r}{v_\mathrm{p}}\right)^2\right)^{-\frac{1}{2}}.
\end{align}
Conversely, for a dust grain trapped at low-$J$ attractor points, the angular velocity can be approximated as the thermal angular velocity, $\omega_\mathrm{th} = \sqrt{2 k T_\mathrm{gas}/I}$ \citep{Hoang2014}. In this case, the ratio becomes:
\begin{align} \label{r_eb_lJ}
    \frac{\abs{\pmb{\Gamma}_\mathrm{ed,max}}}{\abs{\pmb{\Gamma}_{b,\mathrm{max}}}} \approx& \num{5.7e5}\hat{\rho}^{1/2}\left (\frac{a}{\SI{0.1}{\mu m}}\right )^{1.5}\left (\frac{T}{\SI{100}{K}}\right )^{-1}\left (\frac{r}{\SI{1}{au}}\right ) \nonumber \\
    &\times \left(1+\left (\frac{\Omega_\odot r}{v_\mathrm{p}}\right)^2\right )^{-\frac{1}{2}}.
\end{align}

Similarly, the ratio between the magnetic dipole torque, $\abs{\pmb{\Gamma}_{b,\mathrm{max}}}$, and the torque induced by the Barnett effect, $\abs{\pmb{\Gamma}_\mathrm{Bar,max}}$, can be derived from Equations \ref{Tau_b} and \ref{Tau_bar}:
\begin{align} \label{r_b_bar}
    \frac{\abs{\pmb{\Gamma}_{b,\mathrm{max}}}}{\abs{\pmb{\Gamma}_\mathrm{Bar,max}}} &\approx \frac{\abs{\pmb{\mu}_b}}{\abs{\pmb{\mu}_\mathrm{Bar}}} \nonumber \\
    &\approx \begin{cases}
     \num{9.4e-5}\left (\frac{V}{\SI{3}{V}}\right )\left (\frac{T_\mathrm{d}}{\SI{15}{K}}\right )\quad &\text{(silicate)},\\
     \num{0.74}\left (\frac{V}{\SI{3}{V}}\right )\left (\frac{T_\mathrm{d}}{\SI{15}{K}}\right )\quad &\text{(carbonaceous)}.
     \end{cases}
\end{align}

Equations \ref{r_eb_hJ}, \ref{r_eb_lJ}, and \ref{r_b_bar} suggest that, for paramagnetic grains at high-$J$ attractor points, the magnetic dipole moment generated by the Barnett effect typically dominates, making $B$-RAT the prevailing alignment mechanism. In contrast, for weakly paramagnetic grains at high-$J$ attractor points, and for grains at low-$J$ attractor points, the electric dipole torque can surpass both the magnetic dipole and the torque induced by the Barnett effect, leading to $E$-RAT as the dominant alignment mechanism. 

\subsubsection{$k$-RAT and relative effects of radiative to electric torques}

\citet{Lazarian2007a} demonstrated that the third component of RAT ($\Gamma_{\mathrm{RAT,e3}}$) induces grain precession around the radiation direction ($\pmb{k}$). When the radiative precession timescale is shorter than the damping timescale, the radiation direction becomes the axis of grain alignment. This alignment mechanism is referred to as $k$-RAT \citealt{hoang_internal_2023}.

in the heliosphere, while the averaging RAT efficiency for spin up the dust grains is approximately $ \left\langle Q_\Gamma\right\rangle \approx 0.4$ for $a > \bar{\lambdaup}/1.8$ (see Equation \ref{RAT_eff}), the third component of the RAT efficiency for grain precession has much smaller value $ Q_\mathrm{\Gamma,e3} \approx 0.01$ \citep{Hoang2014}. Therefore, the third component of the torque $\Gamma_{\mathrm{RAT,e3}}$ can be expressed as:
\begin{align}
    \Gamma_{\mathrm{RAT,e3}} &= \num{2.16e-29}a^{2}_{-5}\gamma_{\mathrm{rad}}U\bar{\lambdaup}_{0.5}\left(\frac{Q_\mathrm{e3}}{0.01}\right)\,\si{.dyne.cm} \nonumber\\
    &= \num{2.1e-21}a^{2}_{-5}\left (\frac{r}{\SI{1}{au}}\right )^{-2}\left(\frac{Q_\mathrm{e3}}{0.01}\right)\,\si{.dyne.cm}, \label{taue3}
\end{align}
where $a_{-5}=a/\num{e-5}\,\si{.cm}$, $\bar{\lambdaup}_{0.5}=\bar{\lambdaup}/\SI{0.5}{\mu m}$, and $Q_\mathrm{e3}$ is the third component of RAT efficiency \citep{Hoang2014}.

The ratio between the third component of RAT torque and the electric dipole torque can be derived by combining Equations \ref{Tau_ed} and \ref{taue3}:
\begin{align} \label{r_e3_ed}
    \frac{\Gamma_{\mathrm{RAT,e3}}}{\abs{\pmb{\Gamma}_\mathrm{ed,max}}} \approx 15 \left (\frac{V}{\SI{3}{V}}\right )^{-1}\left (\frac{r}{\SI{1}{au}}\right )^{-3}\left(\frac{\SI{e-5}{G}}{B_r}\right)\left(\frac{Q_\mathrm{e3}}{0.01}\right).
\end{align}

Assuming $Q_\mathrm{e3}=0.01$  and referring to the conclusions in Section \ref{mtoe}, Equation \ref{r_e3_ed} suggests that for weakly paramagnetic grains at high-$J$ attractor points, as well as for grains at low-$J$ attractor points, $k$-RAT dominates over $E$-RAT at heliocentric distances $r < \SI{5}{au}$.


\bsp	
\label{lastpage}
\end{document}